\documentclass[acmsmall,screen,nonacm]{acmart}
\setcitestyle{authoryear,open={(},close={)}}
\usepackage{enumitem}
\usepackage{listings}
\usepackage{mathtools}
\usepackage{array}

\newtheorem{theorem}{Theorem}
\newtheorem{lemma}{Lemma}
\theoremstyle{definition}
\newtheorem{definition}{Definition}

\lstdefinestyle{gpecode}{
  basicstyle=\ttfamily\footnotesize,
  columns=fullflexible,
  keepspaces=true,
  showstringspaces=false,
  frame=single,
  framerule=0.2pt,
  breaklines=true,
  tabsize=2
}
\newcommand{\Known}{\ensuremath{\mathsf{Known}}}
\newcommand{\Unknown}{\ensuremath{\mathsf{Unknown}}}
\newcommand{\Clean}{\ensuremath{\mathsf{Clean}}}
\newcommand{\Invalidating}{\ensuremath{\mathsf{Invalidating}}}
\newcommand{\Safe}{\ensuremath{\mathsf{Safe}}}
\newcommand{\Resolve}{\ensuremath{\mathsf{resolve}}}
\newcommand{\ProducerSound}{\ensuremath{\mathsf{ProducerSound}}}

\newcommand{\Normalize}{\ensuremath{\mathsf{normalize}}}

\title{Build-Authorized Evidence for Opaque Calls:
  A Fail-Closed Rewrite-Authority Boundary}

\author{Zhonghua Yi}
\orcid{0009-0006-1721-5843}
\affiliation{%
  \institution{Toka Language Research Group}
  \country{China}
}
\email{yizhonghua@ustc.edu}
\authorsaddresses{Zhonghua Yi, Toka Language Research Group, China;
  \href{mailto:yizhonghua@ustc.edu}{yizhonghua@ustc.edu}}

\def\gpeIncludePublicAppendix{1}
\begin{abstract}
Detached semantic facts about an opaque native provider do not by themselves
justify a compiler rewrite.  Authority must be restricted to the accepted fact,
selected provider and build, actual caller, relevant callback environment, and
exact observation.  The reviewed systems provide library annotations,
summaries, compiler assumptions, contract carriers, and build attestations in
close combinations, but we found no evaluated realization of the residual
integration below.  We use one-hop topology-load reuse as a minimal observable witness of rewrite authority, not
as the paper's optimization target: authorization removes one reload, while
rejection has an explicit neutral behavior.

We present a build-authorized path-effect interface for that deployment
boundary.  An authorization-receipt/link-receipt chain binds an accepted
detached artifact to the unoptimized caller IR, selected provider snapshot,
optimizer entry, caller object, and link inputs; the trusted driver passes
accepted facts through a narrow internal LLVM API.  The design separates three
non-interchangeable authorization failures: receipt closure rejects the wrong
provider or build inputs, callback-environment closure accounts for
independently supplied code, and projection identity confines authority to the
exact selected field.  A
conservative consumer reuses one-hop pointer observations from a
\texttt{noalias} root or one constant nonzero projection.  Rocq proves
conditional refinement and authority non-amplification in bounded abstract
models; applying them to LLVM additionally assumes asserted-effect soundness,
alias admission, and compiler/ABI adequacy.  Dynamic-provider routes also
assume that the protected call resolves at execution to the receipt-bound
provider.
As one producer instantiation, we select Toka for its checked distinction
between payload access and root-topology mutation, ordinary LLVM/native
output, and exportable function/root evidence.  A frozen build supplies a
conservative source-summary gate and exact unoptimized LLVM~20 IR.  A separately
implemented exact-IR checker
scans every instruction in the selected function and accepts only a bounded
topology-preserving subset; only that exact accepted IR is passed to a fixed
\texttt{llc} invocation to produce the receipt-bound final provider object.
Its decision is source-summary-independent: it does not use the source-summary
verdict or authorization metadata as evidence.  This does not imply
organizational or third-party independence; the checker remains trusted.
Rocq proves that acceptance of a bounded action list implies topology-store
identity and hence the existing $\ProducerSound$ obligation.  LLVM parsing,
correspondence between the C++ checker and that action list, and semantic
preservation by the fixed backend remain external premises.
For one Darwin/arm64 embedded-object profile instantiated by three final
images, a post-link checker additionally attributes each protected definition
to its staged provider, excludes a protected dynamic import, decodes the
caller's direct branch target, and binds those facts and the final executable
into a second receipt.  This closes the runtime-target premise for one
protected direct call edge per image, not for the dynamic native-library routes
or a general loader.

Receipt controls leave provider substitution and mismatched build inputs
unauthorized.  Across three issuer-declared real native boundaries, with one
frozen fixture per boundary---prebuilt Darwin \texttt{readv} and
\texttt{recvmsg}, and a Homebrew Cairo projection from a second positive native
library---authorized caller IR retains the opaque call, reduces the relevant
loads from two to one, and preserves the observed results.  A real
\texttt{jpeg\_read\_header} probe is
rejected because its callback environment remains open, whereas an
independently bound synthetic callback singleton demonstrates the supported
closure rule.  The receipt-authenticated, issuer-declared Cairo stable-slot
effect is accepted while a
synthetic distinct-offset projection control and unsupported GEPs remain
neutral.  A receipt-gated lowering to LLVM's existing
\texttt{invariant.group} mechanism matches the custom consumer on the
registered straight-line projection, while a real Clang
\texttt{noescape} redeclaration leaves both loads.  Graded coordinates and
run-length exact paths produce identical code in their common fragment.  The
synthetic checked-producer route consumes three Toka provider objects produced
from exact IR accepted by that checker, retains each call, and reduces two
loads to one.
The contribution is
the checked
deployment--compiler boundary and its explicit trust/applicability frontier,
not a uniquely expressive effect encoding or load-elimination algorithm.
\end{abstract}



\begin{document}
\maketitle

\section{Introduction}
\label{sec:intro}

An effect fact about an opaque native provider is not yet authority to rewrite
a particular caller.  A compiler-facing mechanism must establish which fact
was accepted, which provider and build it names, whether caller-supplied code
extends the effect, and which observation may be reused.  This authorization
problem is especially sharp at a stable native application binary interface
(ABI): the provider body may be in a prebuilt dynamic library, importing
bitcode may be impossible, and replacing the interface changes the program's
public contract.

We use one-hop topology-load reuse as a minimal observable witness of this
authority boundary, rather than as the optimization target of the paper.  A
callee that rewrites pointer topology makes reuse incorrect; one that only
mutates descendant payload permits it; rejection has a concrete neutral
outcome.  The following consumer illustrates that boundary:

\begin{lstlisting}[language=C]
int consume(struct iovec *restrict root, int fd) {
  void *cached = root->iov_base;
  ssize_t n = readv(fd, root, 1);
  if (n != 1 || root->iov_base != cached) return -1;
  return *(unsigned char *)cached;
}
\end{lstlisting}

The public signature of \texttt{readv} exposes that the \texttt{iovec} array
is read through a \texttt{const} pointer, while the operation writes through
the descendant \texttt{iov\_base} buffer.  The declaration alone does not give
LLVM the root-relative Mod/Ref fact needed to establish stability of the
\texttt{iov\_base} slot across the call.  Stock LLVM therefore retains its
second load when the provider body is unavailable.  In the evaluated order-two
model, that slot is the selected inner topology layer; resolving it depends on
the outer root link, whereas writes through its pointer reach descendant
payload.

Precompiled summaries transport effects, compiler metadata carries trusted
assumptions, build attestations bind materials, and callback metadata records
argument forwarding; all are prior art
\citep{guyer1999broadway,guyer2005broadway,rountev2001,li2010,le2005,
llvmlangref2026,clangattributes2026,torresarias2019}.  A detached effect alone is nevertheless
insufficient: the build may select a different provider, caller-supplied code
may extend the effect, and authority for one observation must not spread to
another.  Among the concrete systems reviewed in
Section~\ref{sec:related}, we found no evaluated path from a detached assertion
about an unchanged, build-selected native provider through a receipt-gated
internal optimizer API and then through applicable callback- and
projection-specific scope checks.  We investigate that residual integration,
not effect transport or a new path language.
Broadway evaluates real PLAPACK applications, and the Jikes RVM attribute
pipeline reports SPECjvm98 speedups; our bounded authorization microfixtures
provide no comparable application- or workload-performance evidence
\citep{guyer2005broadway,le2005}.

One non-amplification principle organizes the design: an accepted effect may
authorize only the receipt-bound provenance, relevant executable environment,
and observation identity named by its evidence.
\emph{Receipt closure} binds the provider, caller, pass, and receipt-listed
link inputs.  We call a callback route closed when every external
implementation that can affect the selected observation is independently
bound.  The implemented rule checks this condition only for one direct,
sequential callback.  A callback-open route therefore additionally uses
\emph{callback-environment closure} to determine which separately asserted
callback effects must be composed, under a truthful-forwarding premise.  A
projected observation additionally uses \emph{projection identity} to restrict
authority to one pointer-valued aggregate slot rather than merely one root.
The LLVM consumer separately requires alias admission as an external safety
gate.  These route-specific obligations are not interchangeable: identity
checks do not prove an issuer assertion true, a provider-only effect does not
summarize an unbound callback, and authority for one field does not extend to
its siblings.
The evaluated provider-substitution, callback-rebinding, and synthetic
distinct-offset controls exercise three failure modes for which evidence in
one dimension cannot discharge the others.

Our interface realizes this factoring as a checked deployment--compiler
transaction.  A trusted integrator, acting as contract issuer, publishes a
small path-effect contract.  In the general route this is a trusted manual
assertion; one additional route combines a conservative
Toka~\cite{yi2026toka} source-summary gate with a separately implemented
exact-IR checker.  The source gate checks the producer language's source-level
effect policy and the provenance inputs later bound by the receipt; it does
not establish topology preservation of the final backend input.  The exact-IR checker instead
classifies the unoptimized LLVM IR used to build the final provider object.  A
validator authenticates
and normalizes the contract, snapshots the provider file selected by the
trusted build resolver, and issues an authenticated authorization receipt over
the effect, unoptimized caller IR, provider snapshot, pass schema/plugin,
toolchain, environment, and command.  After code generation, a link receipt
additionally binds the caller object and revalidated link inputs.  The contract
remains detached and the prebuilt provider binary is not rewritten.  Our
custom LLVM 20 passes accept effects only through the driver's narrow internal
API; the public pipeline cannot trigger the rewrite.
The consumers preserve pointer-valued one-hop observations either directly
from a \texttt{noalias} root or through one receipt-bound, constant, nonzero
aggregate projection.  The link gate revalidates the receipt and sidecar and
links staged snapshot bytes.  Rejection withholds rewrite authority: unsupported
IR remains neutral, while receipt or link-input failure may abort the build.
For one deliberately static Darwin/arm64 deployment profile, a post-link
checker also verifies ownership of the embedded provider definition, absence
of a protected import, and the direct branch target, then binds the final image
and certificate into a second receipt.

\paragraph{Contributions.}
This paper makes three bounded claims:

\begin{enumerate}[leftmargin=*]
  \item We factor opaque-call rewrite authority into receipt closure and the
        applicable callback-environment and projection-identity gates.  Their
        fail-closed rules prevent evidence in one dimension from discharging
        another; authentication establishes provenance and scope, while
        semantic truth remains an issuer obligation for the general/manual
        route.
  \item We give conservative LLVM New-PM consumers and mechanize conditional
        refinement, authorization non-amplification, neutral fallback,
        singleton callback closure, and selected-projection separation for
        bounded sequential fragments.  We additionally instantiate the
        producer premise for one bounded Toka route: a source-summary gate
        rejects disallowed language effects, an exact-IR checker accepts only
        a bounded topology-preserving subset, and a fixed
        \texttt{llc} compiles only that accepted IR to the final receipt-bound
        object.  The corresponding Rocq action-list model
        establishes acceptance-to-topology-identity-to-$\ProducerSound$.  A
        bounded post-link profile closes the selected-runtime premise for one
        protected direct edge per image in three embedded-provider
        executables; dynamic-library routes retain that premise.
  \item We evaluate stock compilation, direct-target API splitting, real
        ThinLTO, exact paths, and graded coordinates at prebuilt
        \texttt{readv} and \texttt{recvmsg} boundaries; an open-callback
        \texttt{libjpeg} rejection and receipt-bound callback singleton; and a
        nonzero Cairo projection from a second positive provider.  The
        matrices expose both the useful points and their applicability
        boundaries.  A protocol ablation lowers the same accepted fact
        to fresh LLVM \texttt{invariant.group} identities and matches the
        custom consumer on the registered straight-line projection; a real
        external \texttt{noescape} annotation is too weak for that reuse.  A
        paired micro-build measurement reports the full current transaction
        cost rather than treating authorization as free.  A synthetic
        cross-language route consumes three frozen Toka provider objects
        produced from exact IR accepted by that checker, while source summary,
        checker certificate, backend, final-object, sidecar, and alias controls
        remain fail closed.  A bounded deterministic IR corpus and a
        checked-producer phase/size measurement expose checker coverage and
        prototype cost without claiming checker correctness or scalability.
\end{enumerate}

\paragraph{Claim boundary.}
Graded coordinates and exact paths are extensionally equal in the evaluated
layer-only consumer, and receipt-gated \texttt{invariant.group} matches the
custom lowering in the straight-line control.  The claim is therefore the
checked authorization boundary, not a unique encoding or load-elimination
algorithm.  The proofs and executable results are bounded and conditional on
effect soundness, alias, and compiler/ABI adequacy.  Linker and runtime-target
adequacy remain premises for the dynamic-library routes; the static profile
checks one embedded-provider executable and call edge but is not general
loader enforcement.  The results establish no workload-scale speedup.  The
bounded C/C++ census found five syntactic reuse shapes among 34 direct calls
but admitted no unchanged client; a separate Rust IR search likewise found no
complete accepted window.  These are bounded applicability results, not
ecosystem-prevalence estimates.  The checked Toka route narrows the manual
premise for one synthetic fragment,
while checker/model, backend, and full-compiler correspondence remain
external.

\section{Validated Path-Effect Interface}
\label{sec:interface}

\subsection{Deployment and threat model}

We target a trusted integrator that has obtained an out-of-scope audit or
specification basis for a fixed provider binary but cannot rewrite it.  Acting
as issuer, that integrator normally makes a manual effect assertion about the
provider build; the bounded checked-producer instantiation below instead
combines conservative object-bound Toka source evidence with direct
instruction-level inspection of the exact unoptimized LLVM IR used to build
the final provider object.  A trusted resolver selects a provider pathname for
the current build.
The validator, controlled driver, registered optimizer entry, link gate, and
hash-based message-authentication-code (HMAC) key form the trusted computing
base.  Detached artifacts, selected pathnames, caller modules, intermediate
receipts, and receipt-listed link inputs may be stale, mismatched, or modified
before their respective gates; code already executing inside the trusted
computing base is not an adversary.

The guarantee is authorization non-amplification: no unaccepted or mismatched
input acquires rewrite authority.  It is not a proof that the asserted effect
is true.  For example, a same-symbol, ABI-compatible replacement that mutates
the selected slot must not inherit an artifact issued for the original
provider.  Snapshot and receipt binding make that mismatch neutral or abort the
controlled build.  Linker definition ownership and the later runtime
resolution of a protected dynamic call to the receipt-bound provider remain
explicit adequacy premises for the dynamic-library routes.  A separate bounded
profile embeds the selected provider object and checks one protected direct
call in the final image; it is not a general loader reference monitor.

\subsection{Authorization factoring and trusted boundary}

\begin{figure}[t]
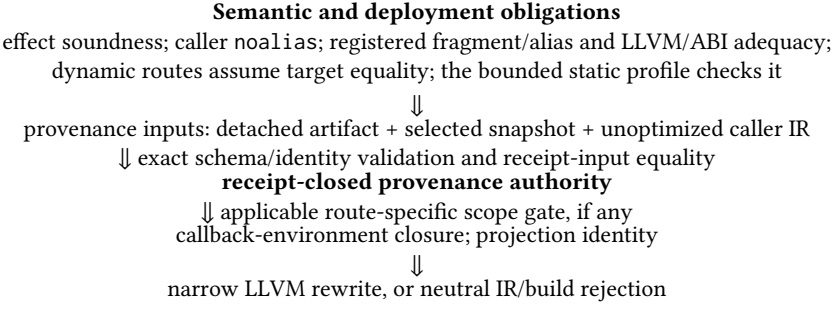

\centering
\small
\begin{tabular}{c}
\textbf{Semantic and deployment obligations}\\
effect soundness; caller \texttt{noalias};
registered fragment/alias and LLVM/ABI adequacy;\\
dynamic routes assume target equality; the bounded static profile checks it\\[2pt]
$\Downarrow$\\[-2pt]
provenance inputs: detached artifact + selected snapshot + unoptimized caller IR\\
$\Downarrow$ exact schema/identity validation and receipt-input equality\\[-2pt]
\textbf{receipt-closed provenance authority}\\
$\Downarrow$ applicable route-specific scope gate, if any\\[-2pt]
callback-environment closure; projection identity\\
$\Downarrow$\\[-2pt]
narrow LLVM rewrite, or neutral IR/build rejection
\end{tabular}
\caption{Authorization flow and trust boundary.  The checked path prevents
registered authority amplification and receipt-input substitution; callback
and projection gates apply only when the call is callback-open or the
observation is projected.
The flow consumes the remaining semantic and compiler-adequacy premises.  The
bounded checked-producer route discharges effect soundness in its abstract
fragment, and the bounded static profile operationally discharges target
equality for one registered call edge.}
\Description{A top-to-bottom authorization flow.  Effect soundness, noalias,
registered semantic-fragment and alias adequacy, compiler adequacy, and target
equality are explicit obligations.  The bounded checked-producer route
discharges effect soundness in its fragment, while the bounded static profile
checks target equality for one registered call edge; dynamic routes retain
that premise.  Detached artifact, selected provider snapshot, and unoptimized
caller IR pass exact validation and receipt equality to produce internal
authority.  Applicable callback or projection checks then permit a narrow LLVM
rewrite, neutral code, or build rejection.}
\label{fig:authorization-flow}
\end{figure}

Figure~\ref{fig:authorization-flow} separates provenance,
executable-environment scope, and observation identity from semantic truth.
Receipt closure is a property of measured build inputs and the internal
optimizer entry.  Callback-environment closure and projection admission refine
the scope of an already authenticated claim.  The general soundness theorem
assumes the asserted effect and alias premises; neither an HMAC nor a provider
digest discharges them.  The checked-producer and static-target subprofiles
separately replace one bounded effect or target premise with executable
evidence; they do not establish the remaining obligations.

\subsection{Base artifact and transaction}

The base layered-effect artifact binds the normalized effect defined below to:

\begin{lstlisting}
callee, ABI digest, semantic fragment,
root argument/order, normalized effect,
issuer, selected-provider digest, HMAC
\end{lstlisting}

The selected provider is read once into a content-bound snapshot.  The
validator checks its digest, sidecar, and identity, then issues an
authorization receipt binding the neutral-module digest, canonical effect, provider
snapshot, sidecar, pass schema/plugin, environment/toolchain, and command
template.  Code generation adds the caller-object digest to a link receipt.
The link gate revalidates every claim and links privately staged caller and
provider bytes.  A successful check supplies the effect out of band to the
internal pass API; rejection supplies no authorization.  Successful validation
binds the contract to the selected snapshot; it does not show that the snapshot
obeys the contract.
The optional static-target profile then checks the final Darwin/arm64 image:
the raw link map must attribute the caller and callee to their staged objects,
the callee must be uniquely defined in executable text with no dynamic import,
and the caller's single AArch64 direct branch must decode to that definition.
A post-link receipt binds the prior build receipt, objects, executable, link
map, canonical certificate, protected symbols, environment, profile, and
checker-tool identities.

\begin{definition}[Base deployment authorization]
Let $\chi$ be a receipt, $I$ the current build inputs, $c$ the validation
context, and $q$ a reuse request.  The base route is summarized by:
\[
\frac{
  \mathsf{CheckReceipt}(\chi)
  \qquad I=\chi.\mathit{inputs}
  \qquad
  \mathsf{BaseAuthorize}(c,\chi.\mathit{candidate},q)
    =\mathsf{Grant}(e)
}{
  \mathsf{DeployBase}(\chi,I,c,q)=\mathsf{Grant}(e)
}.
\]
The conjuncts of $\mathsf{BaseAuthorize}$ are artifact authentication and
binding; surface and normalized-effect well-formedness; exact
root/order/provider request equality; caller alias admission; and
$\Safe(q.\mathit{layer},e)$.  Failure of any premise yields
$\mathsf{Neutral}$ in the abstract semantics; the controlled driver may
instead abort before emitting a binary after a receipt or link-input mismatch.
Callback and projection routes add their respective closure predicates before
the same grant.
\end{definition}

\subsection{Bounded checked-producer instantiation}

\paragraph{Producer selection.}
We sought a concrete producer rather than designing a paper-specific source
language.  The selection criteria were conjunctive: a compiler-checked,
root-sensitive distinction between payload access and topology mutation;
ordinary LLVM/native output without changing the caller ABI; and exportable
per-function/root evidence that can be bound to concrete build products.
Representative ownership, general-effect, and verification/specification
toolchains satisfy different subsets of these criteria
(Section~\ref{sec:related}).  Among the concrete systems examined, Toka was the
closest direct match for our bounded fragment: its compiler classifies
root-sensitive effects, emits the required native artifacts, and exports
object-bound function/root evidence \citep{yi2026toka,toka_language}.  We use
it as an existence witness for the producer interface, not as a claim that
this capability is unique to Toka.

The general artifact interface does not prescribe how an issuer establishes
effect truth.  Our checked-producer instantiation uses two distinct,
conjunctive gates.
First, the frozen Toka compiler produces a preliminary native object and its
object-bound \texttt{.tke}.  The source-summary gate requires exactly one
selected function and root, empty function effects, and root effects contained
in \{\textsf{Read}, \textsf{Write}\}.  \textsf{Rebind},
\textsf{Invalidate}, every other root bit, any function effect, malformed
evidence, or preliminary-object/evidence disagreement yields no artifact.
This gate restricts source-level producer eligibility and validates the
compilation-identity inputs later bound by the receipt; it is not evidence that
the exact backend IR preserves the selected topology.

Second, the same registered compiler emits exact unoptimized LLVM~20 IR.  An
exact-IR checker scans every block and instruction of the selected function
under a request fixing the function, root argument, and root order; it
recomputes the target triple, DataLayout, and IR digest from those IR bytes.

\paragraph{Meaning of source-summary-independent.}
Here and throughout, \emph{source-summary-independent} denotes implementation
and input separation, not third-party or organizational independence.  The
checker is a separate C++ executable that recomputes its decision from the
exact IR and registered request.  It takes neither the \texttt{.tke},
source-summary verdict, GPE sidecar, consumer result, nor an expected
acceptance verdict as an authorization input; favorable LLVM metadata cannot
authorize the route.  The checker is nevertheless shipped in the same
artifact and remains part of the trusted build toolchain.  Its correctness and
correspondence to the Rocq action classifier are external premises.

The checker rejects topology-slot writes, unknown or indirect calls, pointer
escape or casts, dynamic GEPs, unknown pointer merges, atomic or volatile
access, and every unclassified opcode.

On acceptance, the checker issues a certificate for that exact IR and request.
Only the accepted IR is passed to the registered fixed \texttt{llc}
invocation, which produces the final provider object named by the GPE
artifact.  A producer receipt binds the source and compiler identity,
preliminary object, \texttt{.tke}, exact IR, checker identity and certificate,
backend identity and flags, final object, and sidecar.  The preliminary object
bound by \texttt{.tke} is distinct from the final \texttt{llc} object.

This route replaces a manual assertion only for a bounded payload fragment.  It
does not prove source-to-IR equivalence, correspondence between the C++ checker
and the Rocq action model, semantic preservation by the LLVM backend, or full
compiler correctness.

\subsection{Effect domain}

For a root of order $K$, let $\mathit{Layer}_K=\{1,\ldots,K\}$.  The normalized
effect domain is:

\[
\begin{array}{rcl}
e &::=& \Unknown
  \mid \Known(R,M,H)\\
H &::=& \Clean \mid \Invalidating .
\end{array}
\]

$R$ and $M$ are read and modified layer sets.  $\Invalidating$ conservatively
covers free, capture, concurrency, untracked alias writes, and behavior
outside the certified fragment.  Let $\sqcup$ be the hazard join:
$\Clean\sqcup\Clean=\Clean$, and every other pair joins to $\Invalidating$.
Composition is union:

\[
\Known(R_1,M_1,H_1)\mathbin{\oplus}\Known(R_2,M_2,H_2)
=\Known(R_1\cup R_2,M_1\cup M_2,H_1\sqcup H_2),
\]
with $\Unknown$ absorbing.
The interface transports $R$ for effect accounting and composition, but the
registered load-reuse decision consults only $M$ and $H$; no optimization
benefit is attributed to the unused read mask.

\subsection{Base surface encodings}

The base layered-effect validator accepts two issuer-facing encodings:

\[
  G(a,j)
  \qquad\text{and}\qquad
  P(a,d),
\]

where $G$ names an absolute layer and $P$ is a run-length exact path.  The
registered domain requires $j\in\mathit{Layer}_K$ and
$d=K-j\in\{0,\ldots,K-1\}$; values that would select layer zero are rejected.
For an order-$K$ chain:

\[
\begin{aligned}
  \Normalize_K(G(a,j))
    &=\Known(\{j+1,\ldots,K\},\{j\},\Clean),\\
  \Normalize_K(P(a,d))
    &=\Normalize_K(G(a,K-d)).
\end{aligned}
\]

This equality is deliberate.  It makes the exact-path route the exact control
for the registered linear-chain fragment and prevents the systems result from
being credited to a favorable wire encoding.

\subsection{Alias admission}

The effect describes mutations reachable from a root; it does not prove that
unrelated caller pointers cannot alias that root.  The current consumer admits
only root arguments carrying LLVM \texttt{noalias}, obtained from a C
\texttt{restrict} parameter in the registered positives.  Missing admission
is a conservative fallback.  This check is necessary but not sufficient for
the reference-model correspondence: registered caller shape/stratification
and the absence of untracked cross-layer aliases remain external, unvalidated
premises.

\subsection{Callback-environment and projection extensions}

The callback singleton closes one declared callback environment by composing a
provider-local effect with the separately validated effect of the actual direct
callback.  Closure authorization
authenticates both component origins and the declared same-root/order map, and
discovers the caller-to-provider direct callback operand in the neutral module.
Truthful provider-to-callback forwarding on that root remains an issuer
assertion and compiler-to-model premise.  A later closure-specific receipt binds that
authorization and every receipt-listed link input.  Missing, mismatched, or
\Unknown{} callback evidence produces no authority.  Component artifacts use a
distinct schema and cannot enter the base validator directly.

For aggregate loads, the parallel projection artifact binds a layout
descriptor, positive DataLayout-normalized byte offset, access width, root and
result address spaces, and a fixed C ABI descriptor.  Expected, offered, and
requested identities must agree exactly, and only a \emph{Stable} claim
authorizes reuse; \Unknown{} and modifying claims do not.  Within the globally
receipt-bound layout/module, the cache key is the root, offset, width, and
root/result address spaces, so authority for one field does not authorize a
same-layout sibling.  Projection artifacts likewise cannot enter the base
validator or consumer.  The implementation accepts only syntactic,
constant, \texttt{inbounds} GEPs; it does not infer general paths.

\section{Semantic Result Spine}
\label{sec:semantics}

The base model uses a single selected root in a sequential, statically layered,
well-stratified, acyclic pointer chain; untracked cross-layer aliases are
outside the base model and its results.  Let $\Resolve_i(\sigma,r)$ be the
address selected at layer $i$ from root $r$ in store $\sigma$.  The layers on
which that selection depends are:

\[
  \mathit{deps}_K(i)=\{i+1,\ldots,K\}.
\]

\begin{definition}[Safe reuse]
\[
\Safe(i,e) \iff
e=\Known(R,M,\Clean)
\land M\cap\mathit{deps}_K(i)=\varnothing .
\]
\end{definition}

\begin{definition}[Effect soundness]
We retain the mechanization's name $\ProducerSound(B,e)$ for the obligation
that the issuer's asserted effect bounds the body's writes:
$e=\Known(R,M,\Clean)$ and every transition of body $B$ changes cells only at
locations whose certified layer belongs to $M$.  Authentication and fragment
checks are separate.
\end{definition}

\begin{definition}[Neutral and optimized observations]
For one execution $B(\sigma,\sigma')$, the neutral observation is
\[
  N_i(B,\sigma,\sigma',r)=\Resolve_i(\sigma',r).
\]
Given an authorization decision $a$, the bounded optimized observation is
\[
  O_i(a,B,\sigma,\sigma',r)=
  \begin{cases}
    \Resolve_i(\sigma,r) & \text{if $a$ authorizes reuse at $i$},\\
    N_i(B,\sigma,\sigma',r) & \text{otherwise.}
  \end{cases}
\]
Thus rejection is semantic fallback, not an optimization error.
\end{definition}

\begin{theorem}[Topology-address preservation]
\label{thm:reuse}
Suppose the chain from $r$ in $\sigma$ is stratified through order $K$,
$\ProducerSound(B,e)$, executing $B$ produces $\sigma'$, and $\Safe(i,e)$.
Then:
\[
  \Resolve_i(\sigma,r)=\Resolve_i(\sigma',r).
\]
Consequently, replacing the post-call resolution with its cached pre-call
address preserves the registered caller observation.
\end{theorem}

The theorem preserves the selected address, not arbitrary payload values below
it.

\begin{lemma}[Composition]
Assume:
\[
\begin{aligned}
  &\ProducerSound(B_1,e_1),\\
  &\ProducerSound(B_2,e_2).
\end{aligned}
\]
Then
\[
\ProducerSound(B_1;B_2,e_1\oplus e_2).
\]
Moreover,
\[
\Safe(i,e_1\oplus e_2)
\iff \Safe(i,e_1)\land\Safe(i,e_2).
\]
\end{lemma}

\begin{theorem}[Validated consumer refinement]
\label{thm:validated-consumer}
Suppose validation accepts an artifact bound to the complete selected
callee/ABI/fragment/provider identity and normalizes it to $e$.  The invoked
body in the model is selected from that complete binding and the caller's
argument map.
The actual observation root is obtained exclusively by looking up the
validated root-argument index in the same map; it is not supplied
independently.  If the executable checker admits the requested coordinate, the
caller supplies the required alias admission,
$\ProducerSound(B,e)$, and $B$ takes $\sigma$ to $\sigma'$, then the optimized
cached observation equals the neutral post-call resolution.  If validation or
authorization rejects, the consumer is definitionally the neutral consumer.
Applying this result to a native dynamic call additionally assumes that linker
resolution and the runtime call target select that same bound body.
\end{theorem}

The remaining results discharge separate dimensions of the authorization
discipline rather than defining one unrestricted calculus.
Theorem~\ref{thm:receipt} establishes transaction non-amplification; singleton
closure establishes whose effects must be composed; and selected projection
establishes how narrowly the resulting authority may be applied.  In every
case, asserted-effect soundness, alias admission, and the relevant
compiler-to-model correspondence remain premises of consumer refinement.

\begin{theorem}[Link-receipt authorization non-amplification]
\label{thm:receipt}
Represent each input bound by the final link receipt---including the neutral
module, canonical effect, provider snapshot, sidecar, pass schema/plugin,
environment, toolchain, command, and caller object---by an abstract content
identifier.  If the deployed consumer obtains an effect $e$, then the receipt
checker accepted, every current identifier equals its receipt claim, and the
existing artifact authorization returned exactly $e$.  Hence $e$ also has the
validated origin described above.  If receipt acceptance fails or any bound
identifier differs, the abstract consumer selects the neutral observation.
\end{theorem}

\begin{theorem}[Receipt-bound selected-runtime refinement]
\label{thm:deployed-selected}
Suppose deployment grants $e$, the actual invoked body equals the body selected
by the registry from the validated binding and caller root map, the selected
chain is stratified, $\ProducerSound(B,e)$, and $B$ takes $\sigma$ to
$\sigma'$.  Then the receipt was accepted, all current receipt inputs match,
the artifact binding equals the expected binding, the observation root and
provider are obtained from that same binding, and the deployed observation
equals neutral post-call resolution.  Equality of the actual and
registry-selected bodies is an explicit runtime-target adequacy premise, not a
conclusion about a native loader.
\end{theorem}

The static-target profile supplies executable evidence for that equality in
one registered Darwin/arm64 profile instantiated by three final images.  In
each image, the receipt-bound provider object owns the embedded definition,
the protected symbol has no dynamic import, and the caller's decoded direct
branch targets that definition.  This instantiates a premise of the theorem;
it is not a mechanized linker or machine-code correctness result and does not
apply to the dynamic-provider routes.

\begin{theorem}[Singleton callback closure]
For one direct, sequential callback on the selected root, an effect returned
by the singleton callback-closure validator (named \textsc{Open1} in the
mechanization) has independently accepted provider and callback origins, the
registered root/order mapping declaration, and equals their effect
composition.  Missing, \Unknown, mismatched, or same-origin callback evidence
returns no effect.  If the provider truthfully forwards the declared root to
that callback and both component effects are truthful, their safe composition
preserves the selected topology address.
\end{theorem}

\begin{theorem}[Authenticated selected projection]
If deployment returns a selected projection effect, then the receipt was
accepted, all abstract build inputs match, and expected, offered, and requested
bindings and projection identities agree.  The identity has positive offset
and width, the claim is stable, and the external executable-admission gate has
accepted the request.
Under the separate asserted-effect-soundness premise for that projection at the
validated callsite root, the cached pre-call slot equals the neutral post-call
slot.  A same-layout, different-offset sibling is not authorized and has a
distinct abstract cache key.  This is an abstract stable-selected-slot result;
LLVM GEP normalization, layout/ABI correspondence, and executable admission
remain external premises.
\end{theorem}

\begin{theorem}[Bounded source-summary producer]
\label{thm:toka-producer}
Consider the abstract Toka fragment generated by skip, payload read, payload
write, and sequencing.  Its topology-store semantics erases payload operations
to identity.  The bridge rejects handle rebind, root invalidation, raw,
unsafe, and unknown commands.  If it returns an effect $e$, then
$e=\Normalize_2(G(0,1))$, the analyzed command contains none of the rejected
effects, and $\ProducerSound(B,e)$ holds for the fragment's abstract body
$B$.  Consequently, under the same stratification and safety premises as
Theorem~\ref{thm:reuse}, the selected outer handle is preserved.
\end{theorem}

This theorem moves effect truth inside the abstract source model for one
fragment; it does not prove that the concrete Toka compiler classifies every
AST or lowers every accepted program according to that model.

\begin{theorem}[Bounded concrete-IR action model]
\label{thm:checked-ir-producer}
For a finite action list, the checker admits only topology traversal, payload
read, payload write, and scalar steps, and rejects topology-write and unknown
actions.  Acceptance implies that the modeled program is identity on the
topology store.  If the checked-IR gate returns an effect $e$, then
\[
  e=\Normalize_2(G(0,1))
  \quad\text{and}\quad
  \ProducerSound(\mathit{checked\_ir\_program}(actions),e).
\]
Under the existing stratification and safety premises, the selected outer
handle is therefore preserved.
\end{theorem}

This theorem concerns the bounded action-list model, not LLVM syntax.  LLVM
parsing and opaque-pointer analysis, correspondence between the C++ checker
and the action list, source-to-IR correspondence, and semantic preservation by
the fixed backend remain external premises.

\paragraph{Mechanized anchors.}
\begin{sloppypar}
The corresponding kernel-checked declarations are
\nolinkurl{Soundness.topology_address_preservation};
\nolinkurl{EndToEnd.deployed_effect_non_amplification} and
\nolinkurl{EndToEnd.deployed_consumer_refines_neutral}, and
\nolinkurl{EndToEnd.deployed_selected_runtime_refines_neutral};
\nolinkurl{OpenCallback.open1_effect_non_amplification} and
\nolinkurl{OpenCallback.open1_sequential_topology_preservation}; and
\nolinkurl{AggregateProjection.deployed_projection_non_amplification},
\nolinkurl{AggregateProjection.deployed_selected_projection_refines_neutral},
and \nolinkurl{AggregateProjection.sibling_projection_not_authorized}; and
\nolinkurl{TokaProducer.bridge_summary_non_amplification},
\nolinkurl{TokaProducer.accepted_toka_summary_implies_producer_sound}, and
\nolinkurl{TokaProducer.checked_toka_fragment_preserves_outer_handle}; and
\nolinkurl{ConcreteIRChecker.accepted_ir_program_is_topology_identity},
\nolinkurl{ConcreteIRChecker.checked_ir_effect_non_amplification},
\nolinkurl{ConcreteIRChecker.accepted_checked_ir_implies_producer_sound}, and
\nolinkurl{ConcreteIRChecker.accepted_checked_ir_preserves_outer_handle}.
\end{sloppypar}

\paragraph{Proof plan.}
The Rocq development inducts over chain resolution.  Its key frame lemma
states that writes outside $\mathit{deps}_K(i)$ preserve every edge traversed
by $\Resolve_i$; composition follows from effect union and sequential
execution.  Boolean well-formedness and safety checks reflect into the
propositional model.  The pure validator then establishes object and coordinate
binding, and the consumer theorem connects successful authorization to
Theorem~\ref{thm:reuse}; all rejection paths select neutral recomputation.
An independent receipt gate then checks exact equality of the abstract
build-input record before it can invoke that existing authorization path;
Theorem~\ref{thm:receipt} follows by inversion of those two boolean gates.
Theorem~\ref{thm:deployed-selected} then composes that result with the
registry-selected-body theorem while retaining actual-target equality as a
premise.
The callback proof reuses effect and asserted-effect composition; discovery of
the actual direct callback, truthful provider forwarding, and its runtime
same-root mapping remain compiler-to-model and issuer-assertion premises.  The
projection proof exposes only the
selected-projection slice, which contains the requested pointer-valued slot at
the validated root, then reuses the existing write-frame definition; it does
not model the pointee heap.  The checked-producer proof inducts over its
bounded source-command syntax: accepted payload operations are identity on the
topology-store quotient, while every topology-changing or unknown constructor
has a rejected summary.  It then reuses the existing producer-soundness and
path-frame results.  The concrete-IR proof separately inducts over a bounded
action list: checker-action acceptance excludes topology writes and unknown
actions, hence establishes topology-store identity, then reuses the existing
$\ProducerSound$ and path-frame results.
Rocq's kernel checker reports these results closed under the global context.
The receipt and artifact authentication decisions are deliberately abstract
and remain separate from $\ProducerSound$, so provenance cannot be confused
with semantic truth.  The model does not establish hashes, HMAC, filesystem
measurement, public-pass non-bypass, or concrete link-gate behavior; its
neutral branch corresponds to fail-closed semantics, while the implementation
may reject the build rather than emit a neutral binary.  Likewise, alias
admission is an explicit boolean premise: this model does not establish LLVM
\texttt{noalias}, GEP/DataLayout/ABI adequacy, callback discovery, or
compiler-to-model, linker-resolution, or runtime-call-target adequacy.

\section{Implementation}
\label{sec:implementation}

The prototype implements validate, optimize, codegen, and link as a controlled
four-stage transaction.

\paragraph{Validator.}
The validator parses the artifact, checks its schema and HMAC, compares
the callee, ABI, fragment, and root coordinates, verifies the build-selected
provider-snapshot digest, and normalizes graded or exact-path syntax.  It then
issues an HMAC-authenticated authorization receipt over the neutral
module, canonical effect, snapshot, sidecar, pass binary/schema, environment,
toolchain versions, and command template.

\paragraph{LLVM consumers.}
The LLVM 20 New-PM consumers compute forward CFG fixed points over simple,
pointer-valued loads.  The base consumer admits a direct \texttt{noalias} root;
the projection consumer additionally recognizes one syntactic constant
\texttt{inbounds} GEP, normalizes its positive byte offset with the module
DataLayout, and keys the cache by root, offset, width, and address spaces.
Dominating observations seed the cache; a safe opaque call preserves only its
registered keys; an unsafe effect kills them; and a later observation can
reseed them.  Meet points retain only facts valid on every predecessor.
Calls, native ABIs, and provider opacity are preserved.  Accepted effects
enter through exported internal C APIs.  The ordinary public pipelines
construct the passes with empty effect tables and ignore public effect
metadata.
Here ``internal'' denotes the trusted driver's protocol, not process
isolation: a principal able to call those exported APIs directly or read or
replace the prototype HMAC key is inside the trusted computing base.
The packaged experiments use public, fixed reproduction keys to exercise the
protocol; secret-key provisioning, isolation, and rotation are not implemented.
The projection route additionally admits only a declaration and call with C
calling convention, non-variadic \texttt{void} result, two address-space-zero
pointer parameters, and none of the registered ABI-changing parameter
attributes.

\paragraph{Closest-mechanism controls.}
The strongest control reuses the Cairo neutral module, provider snapshot,
canonical effect, and receipt protocol.  It accepts the same provider/effect
evidence through a separately receipt-bound consumer
entry, creates a fresh \texttt{llvm.launder.invariant.group} identity for one
same-basic-block pre-call/post-call selected-load window, applies stock GVN,
and removes the temporary intrinsics and metadata before returning IR.
All receipt claims other than the necessarily distinct consumer entry/schema
are equal.  Pre-existing invariance assumptions, intervening stores, atomics,
fences, unknown calls, or any ordinary projection mismatch reject the route.
This is deliberately a closest implementation baseline, not an unauthenticated
annotation route.  Separately, a real external Clang \texttt{noescape}
redeclaration lowers to \texttt{nocapture} but does not state that the selected
pointee slot is stable.  It is an independent user-responsible expressiveness
control, not a consequence of the path-effect receipt.

\paragraph{Callback closure resolver.}
The singleton resolver independently validates provider-local and
whole-callback artifacts and receipts, verifies distinct origins and the
declared same-root/order map, and asks LLVM to canonicalize the neutral module
before checking that it contains exactly one direct provider call with the
bound callback operand.  It does not inspect the opaque provider body to prove
that the provider forwards that root to the callback; truthful forwarding is
an issuer-assertion/adequacy premise.  Only then does it pass the composed effect to the
internal consumer.  Multiple or indirect calls remain unsupported.
Neither callback-component nor projection artifacts can be parsed as a base
layered-effect artifact; only their dedicated resolver/validator may issue the
internal authorization consumed by the corresponding pass.

\paragraph{Toka exact-IR checked producer.}
The source-summary gate first validates the frozen compiler, source,
preliminary object, object-bound \texttt{.tke}, selected function, and single
root.  A separate LLVM~20 checker then parses and scans the complete exact
unoptimized target function.  It classifies bounded root traversals and scalar
payload operations and rejects topology-slot writes, unknown or indirect
calls, pointer escape or casts, dynamic addressing, unknown pointer merges,
atomic or volatile access, and unsupported instructions.  Metadata and
producer markers cannot by themselves authorize the route.  This checker does
not consume the \texttt{.tke}, source-gate verdict, sidecar, or consumer
result; ``source-summary-independent'' refers only to that input and
implementation separation, not to a third-party trust domain.  On success the checker emits a
certificate over the exact IR digest and request; the
registered fixed \texttt{llc} invocation takes that same accepted IR as its
sole provider-code input.

The source gate is therefore an eligibility screen over language-level effects
and compilation-identity inputs, whereas the route admits the bounded
topology-preservation claim only after the exact-IR checker classifies the
backend input.  The two checks are conjunctive but do not prove the same
property.

An authenticated producer receipt binds source, compiler, preliminary object,
\texttt{.tke}, exact IR, checker and certificate, backend and flags, final
object, and sidecar.  The GPE artifact names the final object, not the
preliminary object.  The final object is then consumed as an opaque provider
by the unchanged controlled C/LLVM consumer path.

\paragraph{Codegen and link gate.}
After optimization, the driver compiles the caller and issues a link receipt
that additionally binds the caller-object digest.  The link entry revalidates
the receipt, sidecar normalization, provider snapshot, pass binary/schema, and
toolchain record.  It writes the receipt-validated caller and provider bytes to
a private staging directory and links those bytes, never reopening the
resolver's original provider pathname.  This closes the build-input
check--use window for the staged caller and provider pathnames at native link;
it makes no claim that every implicit linker/toolchain input is snapshotted.
The registered command contains no competing definition, but the prototype
does not inspect relocation ownership for the dynamic-library routes and
assumes their protected symbols resolve to the staged provider.

The bounded static-target profile permits no additional link arguments and
embeds the receipt-bound provider object in a Darwin/arm64 executable.  A
separate post-link checker parses a raw link map and fixed LLVM~20
\texttt{llvm-nm}/\texttt{llvm-objdump} outputs.  It requires strong text
definitions owned by the staged caller/provider objects, no protected
bind/lazy-bind/weak-bind/chained-fixup/indirect-symbol entry, and exactly one
symbolic direct branch in the caller.  It independently decodes the AArch64
\texttt{BL} immediate and compares its target with the final callee address.
A second HMAC receipt binds this certificate, the image and raw map, the
protected symbols, previous receipt and objects, frozen environment, profile,
and tool digests/versions.  Execution rechecks the receipt and reproduces the
certificate.  The parser, link-map production, disassembler, and linker remain
trusted; this is one static call-edge profile, not general loader integrity.

\paragraph{Fail-closed rewrite authorization.}
When the controlled driver is used, missing, malformed, forged,
provider-mismatched, coordinate-mismatched, topology-invalidating, and unknown
effects leave the neutral IR unchanged.  The same holds for local callee
shadowing, alias-unsafe inputs, unsupported observation shapes, and direct
metadata forgeries.  Receipt, neutral module, sidecar, snapshot, plugin/schema,
caller,
environment, and link-command tampering are rejected before their respective
stages.  Darwin's ABI spelling
\texttt{\textbackslash 01\_readv} is normalized only for effect lookup; the
call symbol itself is not rewritten.

\paragraph{Contract-authoring modes.}
The real-boundary results use trusted manual contracts; the synthetic
exact-IR-checked Toka route is the sole checked-producer instantiation.  For the registered
\texttt{readv} graded/exact-path comparison, ten source tokens and a 332-byte
public manifest are charged.  Those numbers are not assigned to the distinct
singleton-callback and selected-projection schemas: the accepted projection
sidecar is 170 bytes and its canonical internal effect is 246 bytes, while
callback-artifact size is not used as a cost claim.  Automatic effect
inference for general native providers is not part of the implementation
claim.  HMAC
is appropriate to the evaluated trusted-build boundary; an independent
third-party issuer would require a public-key deployment not implemented here.

\section{Evaluation}
\label{sec:evaluation}

One-hop load reuse is the minimal observable witness used to test whether
rewrite authority is granted, withheld, or over-amplified; it is not evaluated
as an end-to-end performance target.  The evaluation asks:

\begin{enumerate}[leftmargin=*]
  \item Does receipt closure prevent unbound build inputs or public metadata
        from acquiring rewrite authority?
  \item Does callback-environment closure distinguish provider-closed calls, a
        real open-callback rejection, and one independently bound singleton?
  \item Does projection identity authorize one nonzero aggregate slot without
        authorizing a sibling or unsupported GEP?
  \item Can a checked source producer instantiate the effect-truth premise for
        one bounded fragment without a manual effect assertion?
  \item When the applicable gates close, can the one-hop witness expose a
        rewrite while the native provider body and source ABI remain opaque?
  \item Can existing LLVM/Clang assumption mechanisms consume the same
        authorized selected-slot fact, and if so, what contribution remains?
  \item What build-time cost does the current controlled transaction add over
        stock code generation and linking of the same microcase?
  \item Can one restricted embedded-provider profile replace the
        runtime-target premise with checked final-image evidence?
\end{enumerate}

\subsection{Evaluation protocol and evidence map}

\paragraph{Protocol.}
All macOS native-library results were rerun on the registered arm64 macOS host
with LLVM 20.1.8 and the exact provider and, where applicable, header digests
recorded in the artifact.  The
strict entry point runs 20 host-independent artifact regressions and exactly
116 registered executable tests with no skips, performs a clean Rocq 9.1.1
build followed by \texttt{coqchk}; \texttt{make paper} independently rebuilds
the manuscript.  Positive authorization probes check neutral and authorized
IR, retained calls, load counts, runtime oracles, and their registered mismatch
controls.  The libjpeg rejection instead checks a neutral route and a legal
callback counterexample.  Claim-specific frozen files record the applicable
tool, host, provider, fixture, and implementation identities; remaining
executable-only controls are covered by the strict run.  These are executable
deployment checks paired with bounded mechanization.
Fixed artifact keys reproduce protocol decisions; they do not evaluate secrecy
against an adversary who holds the key.

\begin{table}[t]
\centering
\small
\caption{Auditable anchors for the bounded claims.}
\label{tab:evidence-map}
\begin{tabular}{@{}
>{\raggedright\arraybackslash}p{0.18\columnwidth}
>{\raggedright\arraybackslash}p{0.35\columnwidth}
>{\raggedright\arraybackslash}p{0.39\columnwidth}@{}}
\toprule
Claim & Formal anchor & Executable anchor \\
\midrule
Artifact/trust &
validated-consumer refinement; receipt-bound selected-runtime refinement &
Single-shot validator controls plus controlled-build tamper/snapshot tests \\
LLVM consumer &
authorized-consumer refinement in the bounded chain model &
Fixed-point, kill, reseed, alias, and binding tests \\
Reuse/compose &
address preservation; asserted-effect/safety composition; kernel-checked by
\texttt{coqchk} &
Single-call, loop, mixed kill/reseed, and wrong-code controls \\
Deployment &
normalization parity; no LLVM adequacy claim &
\texttt{readv}, \texttt{recvmsg}, and \texttt{libjpeg} evidence \\
Callback closure &
singleton non-amplification and sequential preservation &
Bound preserving/invalidating/unknown callback tests \\
Aggregate projection &
selected-slot non-amplification/refinement and sibling separation &
Cairo nonzero-GEP, ABI, and receipt controls plus a synthetic
distinct-offset projection \\
Checked producer &
bounded action-list acceptance, topology identity, producer soundness, and
outer-handle preservation &
Three exact-IR-checked Toka positives; 14 handwritten rejection classes;
17 deterministic IR variants; source-summary, substitution, and alias controls \\
Static target &
instantiates selected-runtime refinement's target-equality premise; no new
linker theorem &
Three embedded-provider images; ownership/import/direct-branch checks and
post-link receipt controls \\
Transaction cost &
no cost theorem claimed &
21 paired stock/authorized builds; 21 checked-producer phase samples and six
checker-size points \\
\bottomrule
\end{tabular}
\end{table}

Section~\ref{sec:semantics} lists the exact kernel-checked declaration names for
the base, receipt, callback, and projection rows.

\begin{table}[t]
\centering
\small
\caption{Receipt-closure controls from the independent synthetic
controlled-build fixture (whose runtime oracle is 40, rather than the real
boundary's 42).  ``Abort'' means the controlled build issues no binary through
its link entry.}
\label{tab:receipt-controls}
\begin{tabular}{@{}p{.43\columnwidth}p{.19\columnwidth}p{.29\columnwidth}@{}}
\toprule
Input or perturbation & Gate & Observation \\
\midrule
All receipt inputs match & optimize/link & loads $2\!\rightarrow\!1$; result 40 \\
Neutral IR, effect, schema, or plugin mismatch & before pass & optimizer not run \\
Receipt, caller, sidecar, snapshot, environment, toolchain, or command tamper &
link & abort \\
Original pathname replaced after snapshot & staged snapshot & result 40 \\
Staged snapshot bytes changed & link & abort \\
Bare public effect metadata & public pass & neutral IR \\
\bottomrule
\end{tabular}
\end{table}

\begin{table*}[t]
\centering
\caption{Registered applicability matrix.  Load counts compare neutral and
authorized caller IR; all calls remain present.  Rejection characterizes the
authorization boundary rather than a failed optimization benchmark.
``Declared'' denotes an issuer assertion accepted and scoped by the validator,
not a fact inferred from the provider body; the Toka row instead uses its
bounded checked-producer route.}
\Description{Six registered boundaries classified by selected observation,
closure and alias premises, authorization result, load count, and the design
boundary each case establishes.}
\label{tab:applicability-matrix}
{\footnotesize
\setlength{\tabcolsep}{3pt}
\renewcommand{\arraystretch}{1.03}
\begin{tabular}{@{}
>{\raggedright\arraybackslash}p{.10\textwidth}
>{\raggedright\arraybackslash}p{.16\textwidth}
>{\raggedright\arraybackslash}p{.20\textwidth}
>{\raggedright\arraybackslash}p{.21\textwidth}
>{\raggedright\arraybackslash}p{.24\textwidth}@{}}
\toprule
Boundary & Selected observation & Closure/admission &
Decision and result & Established boundary \\
\midrule
\texttt{readv}
(\texttt{libsystem}) &
\texttt{iovec.iov\_base}, direct-root lowering &
Declared provider-closed; \texttt{restrict}/\texttt{noalias} root &
Both encodings accepted; loads $2\!\rightarrow\!1$; result 42 &
Fixed ABI and opaque provider; both encodings are tied \\
\texttt{recvmsg}
(\texttt{libsystem}) &
\texttt{msghdr.msg\_name}, direct-root lowering &
Declared provider-closed; root \texttt{noalias} and nested-output
separation &
Both accepted in two alias modes; loads $2\!\rightarrow\!1$; result 42 &
Selected pointer slot is reusable while sibling
\texttt{msg\_namelen} changes \\
\texttt{jpeg\_read\_}\\
\texttt{header}
(\texttt{libjpeg}) &
\texttt{cinfo->err} &
Callback-open: caller \texttt{init\_source} receives the root &
No provider-only contract; loads remain 2; legal callback rebinds the slot &
Provider identity alone does not close caller-supplied behavior \\
\textsf{Callback singleton}
(independent objects) &
\texttt{*root} &
One direct, sequential, same-root callback &
Preserving: $2\!\rightarrow\!1$, result 11;
invalidating/other: 2 loads &
One bound callback; general callback sets unsupported \\
\texttt{cairo\_}\\
\texttt{append\_}\\
\texttt{path}
(\texttt{libcairo}) &
\nolinkurl{cairo_path_t.data}, byte offset 8 &
Declared provider-closed; fixed arm64 layout/ABI and root
\texttt{noalias} &
Declared \textsf{Stable}; validator scopes exactly:
$2\!\rightarrow\!1$; controls retain 2 &
Second positive native library and nonzero projection; synthetic offset-16
separation \\
\textsf{IR-checked Toka}
(three frozen objects) &
outer handle of a synthetic C \texttt{int***} root &
Source summary plus exact-IR certificate; fixed \texttt{llc}; C
\texttt{noalias} root &
All three accepted; loads $2\!\rightarrow\!1$; results 34, 28, and 46;
noalias-free: 2; static subprofile closes the protected direct target &
Checked producer instantiation and one target-closed deployment profile;
synthetic cross-language route, not an unchanged real client \\
\bottomrule
\end{tabular}}
\end{table*}

\subsection{Mechanism and recurrence}

The single-call probe removes one redundant outer load only for the safe
coordinate.  The loop probe removes the loop-carried topology-preserving
reload.  The mixed preserving/invalidating chain removes only the preserving
reload; deliberately reusing through the invalidating effect changes the
oracle result and is rejected by the real consumer.

An alias control stores a descendant pointer that points back to the observed
root slot.  Without \texttt{noalias}, the consumer remains neutral and retains
both loads.  Merely adding \texttt{noalias} would not turn this control into a
defined counterexample: LLVM's ``based on'' rules do not make a pointer loaded
through an argument based on that argument, while parameter \texttt{noalias}
makes it undefined to access a modified location through both based-on and
non-based-on pointers~\citep{llvmlangref2026}.  This control therefore checks
fail-closed behavior when alias admission is absent; broader
LLVM-to-model alias adequacy remains an explicit premise rather than a proved
consequence of the attribute.

\subsection{Exact-IR-checked Toka producer}

The producer uses the public Toka repository~\cite{toka_language}.  We froze
commit \texttt{8056dabc722d}, its compiler executable,
three accepted source inputs, their preliminary native objects and
object-bound \texttt{.tke} summaries, the exact unoptimized LLVM~20 IR emitted
for each input, an exact-IR checker and certificate, a fixed
\texttt{llc} invocation, the resulting final objects, and the
bridge/consumer implementations.  The source-summary gate first requires an
empty function effect and a sole root whose effect is contained in
\textsf{Read}$\mid$\textsf{Write}.  The exact-IR checker then scans every
instruction in the selected function without consulting that summary
decision.  Only accepted IR is
supplied to the fixed backend, and the final object---not the preliminary Toka
object---is named by the issued $G(0,1)$ artifact.

\begin{table}[t]
\centering
\small
\caption{Accepted exact-IR producer cases.  Counts are
(root traversals, payload reads, payload writes); every optimized caller
retains the opaque call.}
\label{tab:toka-producer}
\begin{tabular}{@{}p{.29\columnwidth}p{.22\columnwidth}p{.39\columnwidth}@{}}
\toprule
Source case & Checker counts & Consumer observation \\
\midrule
payload read/write & $(2,1,1)$ &
loads $2\!\rightarrow\!1$; result 34 \\
payload read only & $(2,1,0)$ &
loads $2\!\rightarrow\!1$; result 28 \\
two sequential payload writes & $(4,2,2)$ &
loads $2\!\rightarrow\!1$; result 46 \\
\bottomrule
\end{tabular}
\end{table}

Four source-summary controls---handle rebind, root consume/invalidate, raw
boundary, and unsafe boundary---are rejected before the IR gate.  Separately,
14 handwritten IR controls are rejected: direct topology store; partial-width
topology overwrite; scalar load from a topology slot; unknown call; indirect
call; pointer return/escape; \texttt{ptrtoint}; \texttt{inttoptr}; dynamic GEP;
pointer \texttt{phi}; pointer \texttt{select}; atomic load; volatile store; and
unsupported opcode.  Malformed IR is a tool error, and favorable metadata does
not authorize unsafe input.  A deterministic bounded corpus additionally
exercises six LLVM-valid preserving variants across CFG, scalar type, and
root-spill shapes and 11 LLVM-valid exact-reason rejecting variants.  Every
decision matches its fixed label; this is regression coverage, not fuzzing or
checker-soundness evidence.

The producer receipt detects substitution of the source, preliminary object,
\texttt{.tke}, exact IR, checker, certificate, backend, backend flags, final
object, sidecar, or receipt itself.  Removing the C caller's
\texttt{restrict}/LLVM \texttt{noalias} admission retains two loads in all
three positive cases, so neither source nor IR gate bypasses the independent
alias gate.  These observations establish a checked-effect-export route for
three synthetic payload-only programs.  They do not establish source-to-IR
equivalence, correspondence between the concrete checker and the Rocq action
model, backend correctness, full Toka compiler correctness, general
source-effect inference, or an admitted unchanged real client.

For one accepted producer, 21 warm-host samples give 533.559\,ms median
construction and 127.138\,ms receipt revalidation (660.697\,ms combined).
A separate six-point accepted-IR response spans 10.879--17.764\,ms for
8--7,169 selected-function instructions.  The supplementary appendix gives
the phase breakdown and measurement boundary; these are descriptive prototype
costs, not scalability results.

\subsection{Static target closure}

We relink the three exact-IR checked providers under a stricter
Darwin/arm64 profile with no user-supplied linker arguments.  Each final Mach-O
contains one strong \texttt{mutate\_inner} text definition attributed by the
raw link map to the privately staged provider object.  The protected symbol
appears in no bind, lazy-bind, weak-bind, chained-fixup import, or indirect
symbol entry.  The caller contains exactly one symbolic direct call; decoding
the AArch64 \texttt{BL} immediate yields the final callee address.  A post-link
receipt binds those facts, the executable and raw map, the prior build receipt
and objects, symbols, environment, profile, and checker-tool identities.

All three images preserve one selected load and return 34, 28, and 46.
Changing any bound post-link material, reassigning the callee's link-map
ownership, adding linker arguments, or importing the provider dynamically
rejects.  In a separate control, a same-name injected dylib runs its
constructor, but the protected direct call still returns 34.  This closes the
runtime-target premise for these registered executable/call-edge pairs.  It
does not certify a general linker, arbitrary machine code, process integrity,
or the dynamic-provider experiments.

\subsection{Prebuilt \texttt{readv} boundary}

The selected provider is the system
\texttt{/usr/lib/system/libsystem\_kernel.dylib}.  Its exported
\texttt{\_readv} symbol and SHA-256 digest are checked, and the digest is
rechecked before link.  Every route below returns 42 under default type-based
alias analysis (TBAA) and \texttt{-fno-strict-aliasing}.

\begin{table*}[t]
\centering
\caption{Final fair-control matrix.  ``Unavailable'' means no provider IR is
available to the consumer; $G$ and $P$ are the graded-coordinate and exact-path
encodings from Section~\ref{sec:interface}.  Deployment/API requirements and
written/serialized costs are reported separately.}
\label{tab:final-matrix}
{\small
\setlength{\tabcolsep}{3pt}
\begin{tabular}{@{}p{.17\textwidth}cp{.17\textwidth}p{.27\textwidth}p{.26\textwidth}@{}}
\toprule
Route & \shortstack{Root\\loads} & ABI / provider IR & Deployment/API requirement &
Written + serialized cost \\
\midrule
Stock \texttt{readv} & 2 & fixed / unavailable & none & none \\
Direct-target \texttt{read} & 0 & changed / unavailable &
replace public \texttt{readv} boundary & not comparable \\
ThinLTO + dylib & 2 & fixed / unavailable & caller bitcode &
no detached contract \\
Exact path $P$ & 1 & fixed / unavailable &
sidecar + registered consumer & 10 tokens; 332-B manifest + 114-B artifact \\
Graded coordinate $G$ & 1 & fixed / unavailable &
sidecar + registered consumer & 10 tokens; 332-B manifest + 114-B artifact \\
\bottomrule
\end{tabular}}
\end{table*}

ThinLTO optimizes the available caller and main bitcode but imports no
\texttt{readv} definition from the prebuilt dylib; the call and both root
loads remain.  The direct-target route reaches zero root loads by replacing
the scatter/gather
\texttt{readv(iovec*)} boundary with the direct-buffer \texttt{read(void*)}
boundary.  The graded-coordinate and exact-path routes preserve the original
source ABI and provider opacity while retaining one load.

\subsection{Aggregate \texttt{recvmsg} boundary}

The second system boundary uses the prebuilt Darwin \texttt{recvmsg} symbol
from the same system provider as \texttt{readv}.  The
candidate observation is the pointer-valued \texttt{msghdr.msg\_name} slot;
the contract says nothing about the aggregate's scalar fields.  A real
\texttt{AF\_UNIX}/\texttt{SOCK\_DGRAM} execution writes the descendant payload
and changes \texttt{msg\_namelen} from 128 to 16 while preserving the
\texttt{msg\_name} pointer.  The fixture separately checks that nested output
buffers do not overlap the observed slot.

\begin{table}[t]
\centering
\caption{\texttt{recvmsg} loads under two alias-analysis modes.  The noalias
control uses an accepted sidecar but removes caller admission.}
\label{tab:recvmsg}
\begin{tabular}{lcccc}
\toprule
Mode & Stock & $G$/$P$ & Noalias control & Runtime \\
\midrule
Default TBAA & 2 & 1/1 & 2 & 42 \\
No strict aliasing & 2 & 1/1 & 2 & 42 \\
\bottomrule
\end{tabular}
\end{table}

Every optimized route retains the opaque call and the sibling update.  The
graded-coordinate and exact-path routes again normalize to identical consumer
IR.  One graded route additionally
runs through the controlled transaction: the provider dylib is snapshotted,
the neutral module and effect are authorized, and the link gate consumes only
the digest-checked snapshot.  These results extend the witness from a const
iovec array to a mutable aggregate, but only under the registered
\texttt{noalias}/nested-output-separation admission.  The committed
machine-readable snapshot records the complete six-cell semantic matrix,
provider hash and install identity, toolchain, host, fixtures, and
implementation hashes without host-timing samples.

\subsection{Open-callback \texttt{libjpeg} rejection}

The libjpeg header reader initially appears to be a third positive: LLVM emits
two direct loads of the decompressor's error-manager slot, and the standard
stdio source manager leaves it unchanged.  However, libjpeg invokes a
caller-installed \texttt{init\_source} callback with the complete decompressor
pointer.  Our legal callback replaces \texttt{cinfo->err}, delegates to the
original callback, and still obtains \texttt{JPEG\_HEADER\_OK} on a
$1{\times}1$ image.  Because the candidate observation changes, we issue no
contract: binding only the libjpeg dylib cannot certify an effect whose truth
also depends on caller-supplied callback code.  Its committed snapshot records
both the standard stable execution and the legal callback rebind that forces
rejection.

\subsection{Receipt-bound singleton callback closure}

A separate synthetic boundary tests the smallest supported open-world rule:
one provider object directly calls one callback object on the same root.  The
provider-local and whole-callback artifacts have independent origins,
receipts, and object snapshots.  The controlled driver canonicalizes the
neutral LLVM module, binds the actual direct callback operand, authenticates
the declared same-root/order map, and authorizes only the composed effect.  It
does not inspect the opaque provider body; the claim that the provider really
forwards that root to the callback remains an issuer-assertion/adequacy
premise.

\begin{table}[t]
\centering
\caption{Receipt-bound singleton callback closure.  Every row retains the
opaque call.}
\label{tab:callback}
\begin{tabular}{lccc}
\toprule
Callback evidence & Authorized & Root loads & Runtime \\
\midrule
independently bound preserving effect & yes & 1 & 11 \\
independently bound invalidating effect & no & 2 & $-1$ \\
missing/\Unknown/wrong/mapping mismatch & no & 2 & 11 \\
\bottomrule
\end{tabular}
\end{table}

The invalidating-effect row is a wrong-code control: unauthorized reuse would
change $-1$ to 11.  A module with multiple matching calls is rejected rather
than lifting one safe callback decision to callee-wide authority.  This is a
singleton closure result, not inference for callback sets, indirect calls,
captures, recursion, or dynamic substitution.

\subsection{Authenticated nonzero Cairo projection}

The final positive selects Homebrew Cairo 1.18.4, a provider distinct from the
Darwin system provider.  The registered root is the second argument of the
append-path call.  We observe its pointer-valued \texttt{data} member, at
registered arm64 byte offset 8.
The artifact and receipt bind the provider/header digests, C and LLVM ABI
descriptors, target DataLayout, layout, offset, width, address spaces, pass,
neutral IR, caller object, toolchain, and commands.
The artifact's \textsf{Stable} label is a trusted selected-slot assertion; the
validator authenticates and scopes it but does not infer it from the Cairo
body.

\begin{table}[t]
\centering
\caption{Authenticated Cairo selected-projection result.  Runtime is identical
in every row.}
\label{tab:cairo}
\begin{tabular}{lccc}
\toprule
Claim & Accepted & Projection loads & Call/GEP retained \\
\midrule
missing & no & 2 & yes/yes \\
stable & yes & 1 & yes/yes \\
\Unknown & no & 2 & yes/yes \\
modifying & no & 2 & yes/yes \\
\bottomrule
\end{tabular}
\end{table}

The runtime confirms that the source path header is unchanged and its 11 path
items are copied into the destination context.  A joint IR fixture eliminates
only the real offset-8 pointer reload; a synthetic pointer-valued projection
at offset 16 retains both loads.  The latter is a cache-identity control, not a
claim that Cairo's scalar \texttt{num\_data} field is pointer-valued.  Offset
zero, dynamic or non-\texttt{inbounds} GEPs, missing \texttt{noalias}, wrong
root, local callee shadowing, DataLayout or ABI mismatch, public metadata, and
receipt-input tampering all fail closed.  The registered projection schema
recognizes this single constant shape rather than solving general access
paths.

\subsection{Protocol ablation: receipt-gated LLVM assumptions}

This ablation preserves the provider/effect/receipt authorization protocol but
replaces the custom projection consumer with LLVM's existing invariance
machinery.  We use the same Cairo neutral module, selected provider, canonical
effect, and receipt protocol.  Consumer identity is itself a receipt input, so
the custom and invariance receipts intentionally differ only in consumer
entry/schema; forcing byte-identical receipts would defeat that binding.

\begin{table}[t]
\centering
\caption{Closest-mechanism control on the selected Cairo projection.  All
authorized routes retain the opaque call; missing, \Unknown, or modifying
effects retain two loads.}
\label{tab:invariance-control}
\begin{tabular}{lcc}
\toprule
Route & Authorized & Projection loads \\
\midrule
Stock neutral IR & no & 2 \\
External Clang \texttt{noescape} & independent assertion & 2 \\
Custom GPE consumer & yes & 1 \\
Receipt-gated \texttt{invariant.group} & yes & 1 \\
\bottomrule
\end{tabular}
\end{table}

The independent, user-responsible external \texttt{noescape} redeclaration is
a real Clang source annotation and emits \texttt{nocapture}; the selected-slot
effect does not imply \texttt{noescape}, and \texttt{nocapture} does not
prohibit mutation of the selected pointer-valued field.  The receipt-gated
invariance route instead lowers the accepted selected-slot fact to a fresh
\texttt{invariant.group} identity and
lets stock GVN obtain the same $2\!\rightarrow\!1$ result as the custom
consumer.  Its output contains no temporary invariance metadata or intrinsics.
Injected pre-existing assumptions are rejected; missing, \Unknown, modifying,
tampered-receipt, and intervening-effect controls remain neutral or fail
closed.  In a joint fixture, the selected offset-8 load is reused while the
offset-16 sibling retains two loads.

The parity result closes any claim that the custom cache algorithm is uniquely
necessary: in this fragment, the consumer is an interchangeable lowering.
The remaining evaluated contribution is the authorization protocol deciding
whether a detached selected-slot fact may introduce either lowering.

\subsection{Full transaction cost}

We separately measure the current projection controlled-build implementation against
stock code generation and linking of the same neutral Cairo fixture.  Pass
plugin compilation, neutral C-to-IR generation, contract/sidecar authoring,
and executable runtime are outside the timed region; the runtime oracle is
checked once outside it.  After three discarded warmups, 21 paired
repetitions alternate which route runs first.  Each authorized sample includes
snapshot and authorization, receipt and input rechecks, the LLVM consumer,
caller/main code generation, final recheck, private provider staging, and
native linking.  The stock sample includes the same caller/main code generation
and native link without that protocol.

\begin{table}[t]
\centering
\caption{Warm-host prototype build-transaction time (ms), 21 paired
repetitions.  P95 uses nearest rank.}
\label{tab:transaction-cost}
\begin{tabular}{lrr}
\toprule
Route or phase & Median & P95 \\
\midrule
Stock total & 117.591 & 119.740 \\
Authorization/snapshot & 97.984 & 100.120 \\
Optimize/recheck/codegen & 205.677 & 208.161 \\
Final recheck/stage/link & 211.222 & 213.666 \\
Authorized total & 514.834 & 520.242 \\
Paired authorized--stock increment & 397.492 & 401.527 \\
\bottomrule
\end{tabular}
\end{table}

The current defensive prototype therefore imposes a visible subsecond cost on
this tiny build rather than obtaining authorization for free.  The result
measures the implementation as shipped---including repeated toolchain probes,
provider/header checks, process startup, and private snapshot staging.  It is
an unamortized cost of the defensive build transaction, not a semantic cost
imposed by the Toka producer route or a runtime cost.  It is not an inherent
lower bound, an amortized build-service design, or evidence about scaling to a
large application.

\subsection{Exact-path control and costs}

For the prebuilt \texttt{readv} case, $G$ and $P$ both use a 114-byte artifact,
16 bytes of canonical effect data, a 624-byte caller object, and 272 bytes of
final text.
Their optimized IR is identical.  In the separately measured SQLite matrix,
five registered repetitions of the four graded/exact rows report
0.485--0.512\,ms median pass time and
0.084--0.092\,ms median validation time.  These host-specific diagnostics do
not establish compile- or link-time superiority.  The parity facts close a
graded representation-performance claim while preserving the
deployment-interface result.

\subsection{Bounded unchanged-client census}

We searched the frozen C/C++ scan units for direct calls to
\texttt{cairo\_append\_path}, \texttt{readv}, and \texttt{recvmsg}, manually
removed declarations, definitions, diagnostics, and comments, and retained
only sites with the same pointer-valued root or projection observed before and
after the opaque call.  We then applied the registered gates in order:
alias/root admission, supported projection shape, and closure of intervening
opaque effects.  Table~\ref{tab:client-audit} reports this bounded targeted
census, not an ecosystem-prevalence estimate.  Five of 34 direct invocations
had the syntactic reuse shape; none satisfied all semantic gates.  In
particular, no candidate was rejected solely because of function arity or
return type, so ABI generalization alone would admit none.

\begin{table*}[t]
\centering
\caption{Bounded unchanged-client census.  ``Calls'' counts direct target-API
invocations in each frozen scan unit; ``shapes'' counts the manually screened
pre-call/post-call pointer-observation pattern.}
\Description{Seven frozen source snapshots with direct target-call counts,
reuse-shape counts, and the decisive reason no unchanged client was admitted.}
\label{tab:client-audit}
{\small
\setlength{\tabcolsep}{5pt}
\begin{tabular}{@{}
>{\raggedright\arraybackslash}p{.21\textwidth}
rr
>{\raggedright\arraybackslash}p{.55\textwidth}@{}}
\toprule
Snapshot & Calls & Shapes & Decisive outcome \\
\midrule
Cairo 1.18.4 & 25 & 1 &
Store/local-root seeding unsupported; an intervening opaque call remains \\
GTK \texttt{a8dd18e} & 3 & 0 &
No repeated selected observation \\
libuv 1.52.1 & 1 & 1 &
Root fails alias admission; a user buffer may overlap the request \\
CPython \texttt{v3.14.0} file snapshot & 1 & 0 &
Pointer remains in SSA; no repeated projection \\
NGINX \texttt{95a24d1} & 3 & 2 &
Store/local-root seeding unsupported; selected projection is offset zero \\
curl \texttt{1128177} & 1 & 1 &
Store/local-root seeding unsupported; selected projection is offset zero \\
libevent \texttt{e1f0335} & 0 & 0 &
No direct \texttt{recvmsg}/\texttt{sendmsg} call \\
\midrule
Total & 34 & 5 & No admitted unchanged client \\
\bottomrule
\end{tabular}}
\end{table*}

The machine-readable census records every direct-call position, every retained
shape and decisive gate, and snapshot identities.  The CPython input is
explicitly a tag-labelled single-file snapshot with a content digest, not a
repository checkout or commit-level freeze.  Its checker validates schema,
row counts, and totals without redownloading or reinterpreting upstream
source.
A separate frozen IR-level search across four unchanged Rust FFI snapshots
found 13 calls satisfying the frozen ABI and \texttt{noalias} prerequisites
but no complete accepted load--call--reload window; the supplementary appendix
reports the bounded search and its cross-version positive control.

\subsection{Answers to the evaluation questions}

Yes: the authorized \texttt{readv}, \texttt{recvmsg}, callback-singleton, and
Cairo routes retain the opaque call and ABI while reducing the selected
topology loads from two to one; no provider body is imported.
Receipt closure prevents every registered stale, substituted, or publicly
injected input from acquiring authority inside the trusted-driver protocol;
depending on the stage, rejection retains neutral IR or aborts the controlled
build.  Provider identity alone is insufficient for a callback-open call, as
the libjpeg counterexample demonstrates, while independently bound preserving
evidence closes the supported direct singleton under the stated forwarding
premise.  Finally, the Cairo result and its distinct-offset control show that
authorization follows the complete projection identity rather than spreading
from a root to a sibling slot.  External \texttt{noescape} is too weak for
that selected-slot reuse, whereas receipt-gated \texttt{invariant.group}
matches the custom consumer in the registered straight-line fragment.  The
remaining distinction is the checked route by which a detached provider
assertion acquires authority to introduce that assumption, not a novel RLE
algorithm.  The measured controlled route costs 514.834\,ms at the median
versus 117.591\,ms for the stock micro-build, a paired median increment of
397.492\,ms; this is descriptive prototype cost rather than a scalability
claim.  Finally, the frozen Toka route combines a conservative source-summary
gate with direct instruction-level inspection of the exact IR supplied to a
fixed backend.  Three payload-only programs acquire the registered effect; four
disallowed source summaries, 14 unsafe IR classes, malformed or substituted
materials, and alias-unsafe callers do not acquire rewrite authority.  Together
with the static subprofile, final-image ownership, import, and branch-target
checks replace the runtime-target premise for those three protected call edges;
the dynamic-library routes retain it.  These cases exercise one
non-amplification principle across receipt-bound provenance, callback
environment, observation identity, and the bounded final target.  The
limitations below enumerate the retained semantic and implementation premises.

\subsection{Limitations}

The experiment is a macOS-specific collection of ABI microcases, not an
application benchmark.  Its warm-host transaction timing excludes plugin
compilation, neutral-IR generation, contract authoring, and runtime, and does
not model amortized or parallel builds.
A bounded audit of Cairo/GTK, libuv/CPython, and NGINX/curl/libevent found no
unchanged client that both exposed the registered reuse window and satisfied
its alias admission (Table~\ref{tab:client-audit}).  We therefore do not claim
workload coverage.  This census measures opportunistic adoption by unchanged
existing clients; it does not measure adoption through checked producers that
export evidence during compilation.
Every authorized rewrite assumes sound producer evidence and caller
\texttt{noalias} admission.  For the manual real-boundary routes, if the
issuer assertion or alias premise is false, an accepted rewrite can be wrong;
receipt and identity checks do not detect or mitigate that semantic failure.
The Toka route obtains its topology-preservation claim from
source-summary-independent inspection of the exact unoptimized IR used by the
fixed backend, not from the compiler summary alone.  This is not third-party
verification: the checker remains in the trusted toolchain.  The Rocq result
nevertheless concerns only a bounded action-list model.  LLVM parsing and
opaque-pointer analysis, correspondence between the C++ checker and that
model, backend semantic preservation, source-to-IR intended-behavior
correspondence, and full Toka compiler correctness remain external.
The base graded theorem additionally assumes sequential execution and a
well-stratified acyclic chain; the singleton callback result assumes one
direct, sequential, same-root callback; and the projection theorem exposes a
selected projection slice containing the requested slot, while the executable
Cairo selected-projection result fixes the registered arm64 layout.  Callback
same-root metadata and the selected-slot \textsf{Stable} label are
authenticated declarations, not facts inferred from opaque provider bodies.
The LLVM
consumers handle only simple pointer-valued one-hop loads directly from a root
or through the registered constant nonzero projection; they do not establish
arbitrary access-path optimization, and the base rewrite does not consume the
transported reference mask.  The controlled entry protects its official
trusted-build transaction from bare metadata and input substitution; it does
not defend against an administrator who can replace the driver, plugin, or
HMAC key.  The experiment does not measure workload runtime, automatic effect
inference for general native providers, concurrency, free/capture, general
callback environments, or arbitrary aggregate aliasing.  The checked Toka
fixture is synthetic and does not change the zero-admission result for
unchanged real clients in Table~\ref{tab:client-audit}.  Toka is not an
exclusive producer requirement: another language or verified toolchain could
instantiate the same interface by exporting equivalently checked, object-bound
evidence; this paper evaluates only the bounded Toka route.  In particular,
the \texttt{recvmsg} and Cairo
contracts and rewrites concern selected pointer slots, not a semantic
certification that their whole aggregates are unmodified.
The evaluated \texttt{invariant.group} control handles only one
same-basic-block selected-load window.  Its parity result is implementation
evidence for that common fragment, not a general equivalence theorem for all
CFGs or access paths.
The dynamic-library links contain no competing symbol definition, but their
driver does not verify relocation ownership or enforce that a later dynamic
call resolves to the receipt-bound provider; both remain explicit adequacy
premises.  The static-target profile closes one different, embedded-object
case.  Its link-map parser, LLVM tool output classification, AArch64
instruction decoding, linker behavior, and final-image integrity remain in the
trusted computing base; it is not general loader enforcement.

\section{Related Work}
\label{sec:related}

\paragraph{Modular and precompiled-library summaries.}
Precise relational and symbolic-heap summaries already support body-independent
instantiation, composition, and alias-sensitive specialization
\citep{yorsh2008,dillig2011}.  More directly, Rountev and Ryder construct
points-to and side-effect summaries for statically or dynamically linked
precompiled libraries independently of unknown clients, store a summary with
the library binary, and reuse it when analyzing clients
\citep{rountev2001}.  Their worst-case auxiliary client model conservatively
accounts for callbacks from a library to unknown client modules.  Thus neither
body-free summaries, precompiled-library Mod information, nor conservative
unknown-callback modeling is new here.  Our boundary instead binds the actual
direct callback object and its independently asserted effect when that
supported route is used, while detaching the provider effect from an unmodified
binary and binding both to the controlled build transaction.

\paragraph{Library-level semantic annotations.}
Broadway attaches separate expert-written annotations to C libraries.  Its
\texttt{on\_entry}/\texttt{on\_exit} clauses describe pointer structures,
while \texttt{access}/\texttt{modify} clauses describe per-routine uses and
defs; a whole-program pointer and dataflow framework then uses those facts to
analyze and optimize library calls
\citep{guyer1999broadway,guyer2005broadway}.  Broadway also demonstrates that
such annotations can be applied without requiring source-level changes to the
library or application, and its PLAPACK study supplies real-application
evidence that our bounded authorization fixtures do not.  Thus external
library semantics, pointer-shape and Mod/Ref annotations, and their use in
compiler transformations are prior art.
Broadway studies an integrated source-to-source compilation architecture,
however, rather than fail-closed authorization of a detached claim against the
build-selected native object, exact caller and optimizer entry, link inputs,
callback environment, and projection identity.  Our residual claim is limited
to that latter authorization transaction and decomposition.

\paragraph{Heap graphs, paths, and effects.}
LLVM Data Structure Analysis constructs context-sensitive per-function heap
graphs and supplies Mod/Ref information to alias and dataflow-optimization
clients \citep{lattner2007}.  Deutsch represents interprocedural aliases with
symbolic access paths \citep{deutsch1994}; S\u{a}lcianu and Rinard compute
parameterized results and parameter-rooted mutation-path regular expressions
\citep{salcianu2005}; and DPJ declares nested region-path
\texttt{reads}/\texttt{writes} effects for modular noninterference checking
\citep{bocchino2009}.  IPSSA also resolves record and singleton-heap loads for
interprocedural optimization \citep{calman2009}.  Accordingly, we claim no
novelty for multi-hop effects, path summaries, or heap-analysis-driven
optimization.  Our graded coordinate is only a sufficient quotient for the
evaluated layer-invariant query; the exact-path control is equally precise and
equally compact in the measured fragment.

\paragraph{Candidate checked producers.}
We compared representative source toolchains against the conjunction required
by our producer instantiation, rather than against effect expressiveness
alone.  Rust supplies ownership-based memory safety
\citep{jung2018rustbelt}; Koka exposes inferred row-polymorphic effects
\citep{leijen2014koka}; Low* verifies low-level memory programs and extracts
them predictably to C \citep{protzenko2017lowstar}; and Frama-C supports
specification-driven analysis and verification of C
\citep{cuoq2012framac}.  These designs cover different parts of ownership,
effect checking, native-code production, and memory specification.  Among the
specific designs reviewed here, we did not find the exact detached, object-bound
function/root evidence consumed by our build transaction.  This is a
selection rationale, not an impossibility claim: any of these or another
toolchain could serve as a producer after adding a suitable conservative
exporter.  We selected Toka because its existing root-sensitive effect
classification, LLVM/native output, and object-bound evidence jointly matched
the bounded experiment.

\paragraph{Transported optimizer metadata and LTO}
Le, Lhot{\'a}k, and Hendren encode Soot-produced side-effect information in
Java class-file attributes and consume it in Jikes RVM for CSE, heap SSA,
redundant-load elimination, and LICM \citep{le2005}.  This directly precedes
summary-driven RLE, and its SPECjvm98 evaluation reports execution-time
speedups for several benchmarks, unlike our non-workload microfixtures.  GCC
IPA passes serialize analysis summaries for LTO, and WPA can consume summaries
without bodies \citep{gccipa2026}.  LLVM ThinLTO
combines bitcode with a compact module-summary index
\citep{llvmthinlto2026}; FatLTO objects can contain both native text and
bitcode \citep{llvmfatlto2026}.  These remain IR-bearing LTO pipelines that
may regenerate or import implementations.  Stock LLVM's
\texttt{memory(\ldots)} classes are coarser than a pointer-hop effect
\citep{llvmlangref2026}.  Our comparison concerns continued calls to a
separately selected native provider, not a general claim that LTO is weaker.

\paragraph{Compiler-local contracts and invariance.}
LLVM already exposes assumption-style invariance mechanisms.
\texttt{!invariant.load} imposes value stability wherever the read locations
are dereferenceable, while experimental \texttt{!invariant.group} ties
equal-value assumptions to the same pointer-operand identity; interval
intrinsics can also delimit invariant byte ranges \citep{llvmlangref2026}.
Our protocol-ablation experiment confirms this possibility:
receipt-gated fresh \texttt{invariant.group} identities plus stock GVN match
the custom consumer on the registered straight-line Cairo projection, whereas
a real external Clang \texttt{noescape} redeclaration leaves both loads.
These mechanisms do not themselves describe a
callee's root-relative path effect or authenticate and bind a detached
contract to the build-selected provider.  Our consumer derives only the
supported pre/post-call reuse after artifact validation, root matching, and alias
admission; rejection leaves neutral IR.

Clang permits stronger \texttt{nonblocking}/\texttt{nonallocating}
constraints to be added by redeclaring external-library functions without
editing their headers.  Separately, its user-responsible \texttt{noescape}
attribute is an optimizer assumption about references derived from the
outermost pointer parameter, not a prohibition on pointee mutation.  Clang's
\texttt{callback} attribute also describes a broker's callback operand and
argument mapping and lowers that structure to LLVM callback metadata for
interprocedural analysis
\citep{clangfunctioneffects2026,clangattributes2026}.  Thus ABI-neutral
external annotations, detached source overlays, trusted optimizer assumptions,
and callback-argument maps are prior art.  The residual combination here is a
machine-readable pointer-hop effect, a provider snapshot and authenticated
build/link receipt, independent binding of the actual callback object and its
effect, out-of-band internal-pass authorization, and the supported LLVM
consumer; the identity checks still do not prove the effect semantically true.

\paragraph{Native object metadata.}
A particularly close deployment predecessor is Hewlett-Packard's
object-file-summary scheme, an industrial patent rather than a peer-reviewed systems evaluation
\citep{li2010}.  It stores summary IR, points-to, and Mod/Ref information in
object files and shared libraries, including nested dereference operators, and
lets a linker return that information to a compiler when source is
unavailable.  SELF likewise adds a backward-compatible metadata section to ELF
for binary analysis and transformation \citep{duvarney2003}.  Accordingly,
native/object-library summary transport is prior art, and we make no
first-architecture claim.  The patent does not, however, report a reproducible
semantic or empirical evaluation of the narrower transaction studied here: a
detached contract applied without rewriting the third-party provider binary,
validated fail-closed for rewrite authorization, bound to the build-selected
native snapshot and caller/link receipt, and consumed by our supported custom
LLVM 20 topology-load optimizer.

\paragraph{Certified and signed code contracts.}
Proof-Carrying Code assigns proof construction to the producer and lets a
receiver validate a property of actual machine code \citep{necula1997};
Model-Carrying Code supplies a concise behavior model for consumer policy
checking \citep{sekar2003}; and Security-by-Contract binds mobile code and a
behavioral contract under a digital signature \citep{dragoni2007}.  We
therefore claim neither generic producer/consumer validation nor
code--contract binding.  Indeed, our authenticated artifact is weaker than a
PCC proof: identity checks establish provenance and freshness, while semantic
truth remains a separately checked or trusted-boundary premise.
The in-toto framework cryptographically verifies software-supply-chain
layouts and signed in-toto link metadata---step attestations over functionaries,
materials, products, commands, and byproducts \citep{torresarias2019}.  Thus signed
build-input attestation is also prior art.  Our narrower question is how an
accepted build attestation gates optimizer authority and is then intersected
with callback-environment closure and projection identity.

\begin{table*}[t]
\centering
\caption{Capabilities of the closest prior art for the residual deployment
boundary.  ``Not reported'' describes only the reviewed source; it is not an
impossibility claim.}
\Description{Comparison of carrier and provider relation, validation or
binding, and demonstrated use and scope for six prior-art families and this
work.}
\label{tab:prior-capabilities}
{\scriptsize
\setlength{\tabcolsep}{3pt}
\renewcommand{\arraystretch}{1.05}
\begin{tabular}{@{}
>{\raggedright\arraybackslash}p{.15\textwidth}
>{\raggedright\arraybackslash}p{.24\textwidth}
>{\raggedright\arraybackslash}p{.24\textwidth}
>{\raggedright\arraybackslash}p{.29\textwidth}@{}}
\toprule
Approach & Carrier / provider relation & Validation / binding &
Demonstrated use and relevant scope \\
\midrule
HP OFSI patent \citep{li2010} &
Co-resident summary IR in native objects, archives, and shared libraries;
nested dereference and Mod/Ref &
Association through the containing artifact; detached build receipt not
reported &
Linker-to-compiler optimization feedback; callback/projection authorization
not reported \\
Precompiled-library summaries \citep{rountev2001} &
Points-to and direct/one-dereference summaries stored with a precompiled
library &
Reusable for unknown clients; authentication and exact selected-file receipt
not reported &
Client points-to and side-effect analysis, including conservative unknown-client
callbacks; evaluated load reuse not reported \\
Broadway annotations
\citep{guyer1999broadway,guyer2005broadway} &
Separate library specification describing pointer structure, uses/defs,
analysis properties, and transformations &
Expert-supplied semantics consumed by a source-to-source compiler;
selected-native-object/build receipt not reported &
Whole-program library-call analysis and PLAPACK specialization;
receipt/callback/projection authorization not reported \\
Jikes side-effect attributes \citep{le2005} &
Side-effect and dependence attributes in Java class files &
Attributes identify bytecode locations; selected native-provider receipt not
reported &
Jikes RVM CSE, heap SSA, RLE, and LICM; current callback/projection gates not
reported \\
LLVM / Clang metadata
\citep{llvmlangref2026,clangfunctioneffects2026,clangattributes2026} &
Caller-IR invariance metadata/intrinsics and external declarations or
attributes &
Trusted compiler assumptions and annotations; a detached provider-snapshot
validator is not reported &
Stock optimization and interprocedural analysis; pointer/range identities and
callback maps, not receipt-bound callee effects \\
in-toto \citep{torresarias2019} &
Signed layouts and link metadata over build materials and products &
Cryptographic verification of steps, functionaries, materials, and products &
Supply-chain policy enforcement; path/callback/projection-gated optimizer
authorization is not reported \\
This work &
Detached sidecar for an unmodified prebuilt native provider binary &
Fail-closed binding to provider snapshot, ABI, caller, plugin, and link-input
receipt; effect truth remains a premise &
LLVM one-hop load reuse for provider-closed direct roots, one bound callback
singleton, and one exact constant projection; rejection remains neutral or
aborts the controlled build \\
\bottomrule
\end{tabular}}
\end{table*}

Table~\ref{tab:prior-capabilities} is attribution and boundary accounting, not
a superiority result; its ``not reported'' entries do not imply that build
attestations, external annotations, or LLVM invariance mechanisms cannot be
composed to realize a related design.

\paragraph{Residual boundary.}
Because prior work supplies summaries, signatures, build attestations,
callback maps, and path identities, often in close combinations, the scientific
increment is instead an evaluated decomposition of opaque-call optimizer
authority into separately checked failure modes in the evaluated fragment.  A
same-symbol provider substitution motivates receipt closure; the libjpeg
callback rebind motivates callback-environment closure; and the distinct-offset
control tests observation identity.  The implementation realizes these
obligations in route-specific validators rather than a unified calculus.  In
the concrete systems reviewed,
we did not find this decomposition evaluated across the separate supported
routes as a controlled
optimize--codegen--link transaction over an unmodified native provider binary
selected for one build, one independently bound callback singleton, and one
exact constant projection.

The current system makes no first-architecture, semantic-proof-of-provider,
dynamic-loader rebinding, or post-link substitution claim.  Linker ownership
and runtime resolution to the receipt-bound provider remain explicit premises
for the dynamic-library routes.  The bounded static-target profile
operationally discharges that premise for its registered embedded-provider
direct call edge, subject to its trusted link-map/tool parsing, instruction
decoding, linker behavior, and final-image integrity.  Its HMAC assumes a
trusted build service, and its novelty claim is limited to this non-amplifying
integration and evaluated trust/deployment frontier.

\section{Conclusion}
\label{sec:conclusion}

Opaque-call optimization is an authorization problem as much as an effect
representation problem.  The common invariant in this work is that rewrite
authority must not spread beyond receipt-bound provenance, an applicable bound
callback environment, or the selected observation identity.  Our prototype
therefore grants bounded topology-load reuse only after closing the relevant
build transaction, accounting for the callback environment that can affect the
root, and matching the exact observation to be reused.  This organization
explains both the positive and negative results: prebuilt \texttt{readv} and
\texttt{recvmsg}
are issuer-declared provider-closed positives; libjpeg remains callback-open; the singleton
validator closes one independently bound callback under a truthful-forwarding
premise; and Cairo's issuer-declared stable-slot effect authorizes one nonzero
projection while a synthetic distinct-offset control prevents authority
conflation.
An LLVM-native control further shows that a receipt-gated lowering to fresh
\texttt{invariant.group} identities can match the custom consumer on the
registered straight-line projection; external \texttt{noescape} cannot.
A checked-producer instantiation additionally combines a conservative Toka
source-summary gate with direct instruction-level inspection of the exact
unoptimized LLVM~20 IR used to build the provider.  This denotes a separate
decision procedure and input path, not third-party independence.  Three
synthetic payload-only programs are accepted, 14 handwritten unsafe IR classes
are rejected, and only accepted IR is compiled by the fixed \texttt{llc} into
the receipt-bound final object consumed by a C/LLVM caller as an opaque
provider.
A bounded post-link profile additionally embeds those three checked providers,
attributes the protected definition to the staged object, excludes a protected
dynamic import, and verifies that the caller's decoded direct branch targets
that definition before issuing a final-image receipt.  It closes the
runtime-target premise for those registered call edges, not for the real
dynamic-library routes or a general loader.

Exact paths match graded coordinates throughout their evaluated layer-only
common fragment.
Accordingly, the contribution is the checked boundary by which a detached fact
acquires narrowly scoped compiler authority while an unmodified native provider
binary and ordinary ABI remain opaque, not a uniquely expressive effect
encoding or a uniquely necessary load-elimination algorithm.
The Rocq development proves conditional non-amplification and refinement for
the bounded base, singleton-callback, selected-projection, and action-list
models.  Concrete effect truth, alias admission, compiler/ABI correspondence,
checker/backend adequacy, and dynamic-library runtime targeting remain
external as detailed in Section~\ref{sec:evaluation}.  The static profile's
linker/tool parsing and direct-target checks are trusted executable evidence,
not a mechanized linker theorem.  General callback sets, arbitrary paths, and
runtime-loader integrity remain outside scope.

\section*{Generative-AI Use Disclosure}

OpenAI Codex was used throughout this project to assist with research
planning, literature triage, drafting and revising prose, and generating,
reviewing, and debugging prototype code, tests, artifact scripts, and Rocq
mechanization.  The human author selected the research claims, inspected and
executed the resulting artifacts, verified cited sources and generated
material, and takes full responsibility for the work.

\section*{Data-Availability Statement}

A reproducibility artifact has been prepared for archival submission.  It
contains all source code, contracts, frozen JSON evidence, LLVM 20 pass
sources, executable test
harnesses, Rocq sources, and manuscript sources; no external research dataset
is used.  The command \texttt{make proof} checks the mechanized fragment
independently.  On a host with Python 3.10 or newer and \texttt{make},
\texttt{make portable-check} verifies the package manifest when present,
Python/JSON syntax, the Rocq forbidden-declaration audit, frozen claim values,
cross-file invariants, and 20 portable artifact-regression tests.  This
includes offline validation of the packaged checked-Toka sources, preliminary
objects and \texttt{.tke} files, exact IR, checker certificates,
fixed-backend final objects, sidecars, producer receipts and canonical claims,
and frozen results; it does not invoke
\texttt{tokac}, the native checker, or the backend, or regenerate the recorded
native evidence.  It also authenticates the frozen static-target caller
objects, final Mach-O images, raw link maps, canonical target certificates,
and post-link receipts without re-running the Darwin target checker.  A strict
run of the
full 116-test registered suite requires the exact native environment because
the evaluated protocol intentionally binds provider and header bytes: arm64
macOS 26.5 (Darwin 25.5.0), LLVM/lld 20.1.8, the recorded system provider,
Cairo 1.18.4, \texttt{jpeg-turbo} 3.1.4.1, and the registered Toka
checkout and compiler identity recorded in the checked-producer evidence.  The
\texttt{TOKA\_REPO} and \texttt{TOKAC} environment variables may select those
external inputs for live replay.  The strict
\texttt{make ae-check} entry point runs the 20 artifact checks, seven
LLVM~22 reporter/LLVM~20 consumer alignment checks, exactly 116 registered
native tests with no skips, and then the Rocq checker.  The
artifact records provider paths,
versions, and digests and reports a mismatch rather than silently substituting
another provider.  Its documentation gives claim-specific commands, expected
observations, and approximate runtimes.  An optional
\texttt{make linux-arm64-check} target rebuilds the three exact-IR checked
producer cases in a digest-pinned Ubuntu/LLVM~20 Linux/ARM64 image, emits fresh
ELF objects and platform-specific receipts, and replays the positive and
fail-closed matrices.  Its frozen JSON records the image, tools, target
triple/DataLayout, outputs, and rejection reasons; it reports no container
performance and does not reproduce the Darwin native-library experiments.
We intend to submit the artifact for artifact
evaluation and, upon acceptance, archive it in a persistent public repository.

\bibliographystyle{ACM-Reference-Format}
\bibliography{references}
\ifdefined\gpeIncludePublicAppendix
  \clearpage
  \appendix
  \section{Authorization and Proof Audit}
\label{app:audit}

This appendix makes the bounded scientific argument auditable without opening
the artifact archive.  It collects the exact authorization routes, the formal
definitions hidden by the paper's higher-level notation, the proof
factorization, the checked-producer instantiation, and the remaining adequacy
ledger.  It deliberately omits wire bytes, proof scripts, test logs, and
reproduction commands: those are replication material rather than missing
parts of the argument.

\subsection{Model and failure outcomes}
\label{app:model}

Let locations be natural numbers.  A topology store is a partial successor
map $\sigma:\mathit{Loc}\to\mathit{option}(\mathit{Loc})$, and a certified
layering is $\ell:\mathit{Loc}\to\mathbb{N}$.  Following $d$ edges is:
\[
\begin{aligned}
  \mathit{walk}(0,\sigma,r) &= \mathsf{Some}(r),\\
  \mathit{walk}(d+1,\sigma,r) &=
  \begin{cases}
    \mathit{walk}(d,\sigma,r') & \sigma(r)=\mathsf{Some}(r'),\\
    \mathsf{None} & \sigma(r)=\mathsf{None}.
  \end{cases}
\end{aligned}
\]
Bottom-up absolute layer $i$ of an order-$K$ root is
$\Resolve_i(\sigma,r)=\mathit{walk}(K-i,\sigma,r)$ when $i\le K$, and
$\mathsf{None}$ otherwise.  A chain is stratified when
\[
  d\le K\land\mathit{walk}(d,\sigma,r)=\mathsf{Some}(l)
  \Longrightarrow \ell(l)=K-d .
\]
Writing only layers $M$ means
\[
  \mathit{writesOnly}_{\ell,M}(\sigma,\sigma')
  \iff
  \forall l.\ \ell(l)\notin M\Longrightarrow\sigma'(l)=\sigma(l).
\]
For a relational body $B$, the exact semantic premise consumed by the proof is
\[
\ProducerSound_\ell(B,\Known(R,M,\Clean))
\iff
\forall\sigma,\sigma'.\
B(\sigma,\sigma')\Longrightarrow
\mathit{writesOnly}_{\ell,M}(\sigma,\sigma'),
\]
and it is false for $\Unknown$ and $\Invalidating$ effects.  Thus artifact
authentication never establishes $\ProducerSound$.

The executable routes below return one of three operational outcomes:
\[
  \mathsf{Grant}(e),\qquad \mathsf{Neutral},\qquad
  \mathsf{Abort}(m).
\]
$\mathsf{Grant}(e)$ makes a canonical effect available only to the internal
optimizer entry.  $\mathsf{Neutral}$ withholds authority and preserves neutral
IR or recomputes the post-call observation.  The abstract Rocq development
maps every rejection to $\mathsf{None}$ and hence to neutral observation; the
controlled driver may instead use $\mathsf{Abort}$ for a receipt or link-input
mismatch and issue no binary.

\subsection{Exact authorization routes}
\label{app:routes}

The following pseudocode is intentionally logical rather than a transcription
of JSON or binary parsers.  A base binding contains schema, callee, ABI,
semantic fragment, root-argument index, order, and selected-provider identity.

\begin{minipage}{\dimexpr\linewidth-10pt\relax}
\begin{lstlisting}
BaseAuthorize(ctx, candidate, request):
  require candidate.surface exists as s
  require candidate.binding = ctx.expected_binding
  require argument(s) = ctx.root_argument
  require surface_wf(ctx.order, s)
  e := normalize(ctx.order, s)
  require effect_wf(ctx.order, e)
  require authenticate(ctx, candidate)
  require request.(root_argument, order, provider)
          = ctx.(root_argument, order, provider)
  require request.alias_ok
  require Safe(request.order, request.layer, e)
  return Grant(e)
\end{lstlisting}
\end{minipage}

Every failed \texttt{require} returns $\mathsf{Neutral}$.  The concrete base
parser additionally checks bounded length, schema, issuer, HMAC, encoding,
callee, ABI digest, provider-implementation digest, and fragment before
normalization.  The LLVM consumer separately admits only a registered direct
declaration and a tracked \texttt{noalias} root in the supported load shape.

A receipt closes the later build transaction around the already checked base
route:

\begin{minipage}{\dimexpr\linewidth-10pt\relax}
\begin{lstlisting}
DeployBase(receipt, current_inputs, ctx, request):
  require CheckReceipt(receipt)
  require current_inputs = receipt.inputs
  return BaseAuthorize(ctx, receipt.candidate, request)
\end{lstlisting}
\end{minipage}

The equality covers receipt schema and phase, neutral module, canonical effect,
provider snapshot, sidecar, pass schema and plugin, environment, toolchain,
arguments, commands, and caller object.  The implementation checks the
relevant prefix when authorization is issued, rechecks it before optimization
and code generation, adds the caller-object digest, and rechecks the complete
link receipt before linking privately staged bytes.  A mismatch in this
controlled transaction is an $\mathsf{Abort}$; it is $\mathsf{Neutral}$ in the
abstract observation semantics.

The supported callback rule closes exactly one separately authenticated,
direct, sequential callback:

\begin{minipage}{\dimexpr\linewidth-10pt\relax}
\begin{lstlisting}
Open1(provider, callback?, actual_symbol, query_layer):
  require CheckProvider(provider)
  require callback exists and CheckCallback(callback)
  require Origin(provider) != Origin(callback)
  require actual_symbol = Symbol(callback)
  require provider.(root_argument, order)
          = callback.(root_argument, order)
  require provider.effect = Some(ep)
  require callback.effect = Some(ec)
  e := ep (+) ec
  require Safe(provider.order, query_layer, e)
  return Grant(e)
\end{lstlisting}
\end{minipage}

Here $(+)$ unions read/modified layers and joins hazards.  Missing or
$\Unknown$ evidence, a same-origin component, symbol or root/order mismatch,
or an unsafe composition returns $\mathsf{Neutral}$.  The executable route is
narrower still: it requires one direct provider call in the neutral module,
extracts its callback operand, freezes the registered argument map, and binds
the resulting closure authorization into its own receipt.  It does not prove
that the opaque provider truthfully forwards the declared root.

For a selected aggregate pointer slot, let
\[
 p=(\mathit{layout},\mathit{offset},\mathit{width},
     \mathit{rootAS},\mathit{resultAS}).
\]
Authorization is:

\begin{minipage}{\dimexpr\linewidth-10pt\relax}
\begin{lstlisting}
AuthorizeProjection(ctx, candidate, request):
  require candidate.identity exists as p
  require AuthenticateProjection(ctx, candidate)
  require candidate.binding = ctx.expected_binding
  require request.binding   = ctx.expected_binding
  require p                = ctx.expected_identity
  require request.identity = ctx.expected_identity
  require p.offset > 0 and p.width > 0
  require request.alias_ok
  require candidate.claim = Stable
  return Grant(p, Known({}, {}, Clean))
\end{lstlisting}
\end{minipage}

The same receipt-acceptance and exact-current-input wrapper surrounds this
route.  The empty effect is interpreted only over the selected
$(p,\mathit{root})$ slot, not as whole-heap purity.  The executable consumer
also requires the registered ABI/DataLayout and a syntactic constant
\texttt{inbounds} GEP.  Within the receipt-bound layout and module, its cache
key contains root, offset, width, and both address spaces; a sibling offset
therefore cannot inherit the grant.

The bounded checked-producer route uses two distinct, conjunctive gates before
constructing a base candidate.  The first gate checks source-derived evidence;
the second checks the exact LLVM IR that will be passed to a fixed backend:

\begin{minipage}{\dimexpr\linewidth-10pt\relax}
\begin{lstlisting}
IssueFromToka(compiler, source, checker, backend, fn, root):
  require compiler.commit = 8056dabc...211824
  (preliminary, evidence) = compiler.emitSummary(source)
  require evidence binds source, compiler, target, and preliminary
  require evidence digest is embedded in preliminary
  require exactly one function matches fn
  require that function contains exactly one root, matching root
  require fn.local_effects = fn.effects = {}
  require root.local_effects, root.effects subset {Read, Write}

  ir = compiler.emitExactLLVM20(source)
  require ir embeds the accepted evidence digest
  request = (fn, root-argument=0, root-order=2,
             target-triple, DataLayout, Hash(ir))
  certificate = checker.check(ir, request)
  require certificate accepts this exact ir and request

  flags = fixedBackendFlags()
  final = backend.compile(ir, flags)
  artifact = IssueBase(G(0,1), callee=fn,
                       provider=Hash(final))
  receipt = Bind(source, compiler, preliminary, evidence,
                 ir, checker, certificate, backend, flags,
                 final, artifact)
  return (final, artifact, receipt)
\end{lstlisting}
\end{minipage}

The preliminary object is used only to bind the source summary; it is
different from the final provider object.  The exact-IR checker scans every
instruction in the selected function and rejects
topology-slot writes, unknown or indirect calls, escape or pointer casts,
dynamic GEPs, unknown pointer merges, atomic or volatile access, and
unclassified instructions.  It is a separate implementation and does not read
the source summary, \texttt{.tke}, sidecar, consumer result, or expected
verdict.  This is not third-party or organizational independence: the checker
remains in the trusted toolchain.
Only accepted exact IR reaches the registered LLVM~20 \texttt{llc}; the
ordinary GPE artifact names the resulting final object.  The producer receipt
binds both chains and every material executable input.  The abbreviated commit
above is the frozen full object
\nolinkurl{8056dabc722d2338ede791b330f6024a45211824}.

\subsection{Proof factorization}
\label{app:proof-factorization}

The mechanized argument is short because each authorization layer removes
authority rather than inventing a new semantic frame theorem.

\begin{enumerate}[leftmargin=*]
  \item \textbf{Path frame.}
  \nolinkurl{Model.walk_preserved} is proved by induction on walk depth: if each
  traversed successor cell is unchanged, the resulting walk is unchanged.
  Stratification turns a depth on the walk into its absolute layer.

  \item \textbf{Semantic preservation.}
  \nolinkurl{Soundness.topology_address_preservation} obtains
  \nolinkurl{writes_only} from $\ProducerSound$, and $\Safe$ shows that every
  traversed layer is outside the modification set.  The path-frame lemma then
  yields
  $\Resolve_i(\sigma,r)=\Resolve_i(\sigma',r)$.
  \nolinkurl{Effects.safe_compose} and
  \nolinkurl{Soundness.producer_sound_compose} lift the same result over
  sequential composition.

  \item \textbf{Base consumer.}
  \nolinkurl{Artifact.validate_sound} reflects successful binding,
  normalization, well-formedness, and the external authentication gate.
  \nolinkurl{Consumer.authorize_sound} adds coordinate equality, alias admission,
  and reflected safety.  The authorized branch invokes semantic preservation;
  \nolinkurl{Consumer.rejected_consumer_is_neutral} makes the other branch
  definitionally neutral.

\item \textbf{Receipt and callback wrappers.}
      \nolinkurl{EndToEnd.deployed_effect_non_amplification} inverts receipt
      acceptance and exact input equality before exposing the base authorization.
      \nolinkurl{EndToEnd.deployed_selected_runtime_refines_neutral} composes
      that result with registry-selected binding, root, provider, producer
      soundness, and consumer refinement while retaining actual-body equality
      as a premise.
      \nolinkurl{OpenCallback.open1_effect_non_amplification} similarly exposes
  both independently checked origins, the actual symbol, root/order mapping,
  component effects, their exact composition, and safety.
  \nolinkurl{OpenCallback.open1_sequential_topology_preservation} then reuses
  the composition and frame results.

  \item \textbf{Projection wrapper.}
  \nolinkurl{AggregateProjection.projection_authorize_sound} reflects exact
  binding and projection identity.  The selected-slot producer premise applies
  the existing frame definition only to that slot.
  \nolinkurl{deployed_selected_projection_refines_neutral} combines receipt
  non-amplification with selected-slot preservation, while
  \nolinkurl{sibling_projection_not_authorized} follows from offset inequality.

  \item \textbf{Checked producer.}
  \nolinkurl{TokaProducer.bridge_summary_non_amplification} reflects the
  payload-only whitelist and exact issued effect.  Induction over the bounded
  source-command syntax proves that every accepted command is identity on the
  topology-store quotient, yielding
  \nolinkurl{accepted_toka_summary_implies_producer_sound}.
  \nolinkurl{checked_toka_fragment_preserves_outer_handle} then reuses the
  existing path-frame theorem for that source model.
  Independently, \nolinkurl{ConcreteIRChecker.accepted_ir_program_is_topology_identity}
  proves topology identity for an accepted list in the bounded concrete-IR
  action model;
  \nolinkurl{checked_ir_effect_non_amplification} reflects the fixed effect,
  \nolinkurl{accepted_checked_ir_implies_producer_sound} derives the abstract
  producer premise, and
  \nolinkurl{accepted_checked_ir_preserves_outer_handle} reuses the existing
  path-frame theorem.  These are theorems about the bounded source and action
  models; correspondence from the C++ checker to the action list is not hidden
  inside either proof.
\end{enumerate}

All named declarations are closed by Rocq's kernel under the development's
global context.  This is a proof of the abstract authorization/refinement
spine, not a proof of cryptography, arbitrary native-provider bodies, Toka
compiler correctness, the LLVM parser or C++ checker, fixed-backend
correctness, or LLVM semantics.

\subsection{Claim, evidence, and adequacy ledger}
\label{app:ledger}

Table~\ref{tab:appendix-ledger} separates kernel-checked implications,
executable observations, and premises supplied outside the model.  The test
families summarize the registered source-level observations; opening the
archive is needed to rerun them, not to determine what each claim depends on.

\begin{table}[H]
\centering
\scriptsize
\setlength{\tabcolsep}{3pt}
\caption{Audit ledger.}
\label{tab:appendix-ledger}
\begin{tabular}{@{}p{.16\columnwidth}p{.29\columnwidth}
                  p{.29\columnwidth}p{.16\columnwidth}@{}}
\toprule
Claim & Kernel-checked anchors & Executable observation & External \\
\midrule
Normalize and compose &
\nolinkurl{exact_path_normalizes_to_graded};
\nolinkurl{safe_compose};
\nolinkurl{producer_sound_compose} &
graded/exact routes produce identical IR in the common fragment; recurrence
and kill/reseed controls &
E1, E2, E5 \\
\addlinespace
Base refinement &
\nolinkurl{walk_preserved};
\nolinkurl{topology_address_preservation};
\nolinkurl{authorize_sound};
\nolinkurl{authorized_consumer_refines_neutral} &
single-shot, CFG fixed-point, wrong-coordinate, alias, local-shadow, and
wrong-code controls &
E1--E7 \\
\addlinespace
Receipt closure &
\nolinkurl{deployed_effect_non_amplification};
\nolinkurl{bound_input_mismatch_is_neutral};
\nolinkurl{deployed_consumer_refines_neutral};
\nolinkurl{deployed_selected_runtime_refines_neutral} &
Tampering with neutral IR, effect, plugin, caller, provider, environment,
toolchain, or command does not reach the registered rewrite &
E1--E7 \\
\addlinespace
Static target &
\nolinkurl{deployed_selected_runtime_refines_neutral} under its selected-body
equality premise &
three embedded-provider images have one owned text definition, no protected
import, and one decoded direct branch to that definition; post-link mutation,
dynamic-provider, and same-name injection controls fail closed &
E1--E5, E7, E10; target-check/linker adequacy \\
\addlinespace
Singleton callback &
\nolinkurl{open1_effect_non_amplification};
\nolinkurl{open1_sequential_topology_preservation} &
preserving singleton rewrites; invalidating, missing, wrong, and mapping
mismatch remain neutral; libjpeg remains open &
E1--E8 \\
\addlinespace
Selected projection &
\nolinkurl{projection_authorize_sound};
\nolinkurl{deployed_projection_non_amplification};
\nolinkurl{deployed_selected_projection_refines_neutral};
\nolinkurl{sibling_projection_not_authorized} &
Cairo offset 8 rewrites; offset 16 sibling, dynamic/non-\texttt{inbounds} GEP,
ABI/layout mismatch, and public metadata remain neutral &
E1, E3--E7, E9 \\
\addlinespace
Checked Toka producer &
\nolinkurl{bridge_summary_non_amplification};
\nolinkurl{accepted_toka_summary_implies_producer_sound};
\nolinkurl{checked_toka_fragment_preserves_outer_handle};
\nolinkurl{accepted_ir_program_is_topology_identity};
\nolinkurl{checked_ir_effect_non_amplification};
\nolinkurl{accepted_checked_ir_implies_producer_sound};
\nolinkurl{accepted_checked_ir_preserves_outer_handle} &
three exact-IR positives certify traversal/read/write counts
$(2,1,1)$, $(2,1,0)$, and $(4,2,2)$, rewrite $2\!\rightarrow\!1$, retain the
call, and return 34, 28, and 46; 14 IR classes and four source-summary classes
are rejected; six deterministic preserving and 11 deterministic rejecting IR
variants match their fixed labels; every material-input mutation fails closed;
and absent \texttt{noalias} remains neutral &
E2--E7, E10 \\
\bottomrule
\end{tabular}
\end{table}

\begin{description}[leftmargin=2.4em,style=nextline,font=\normalfont\bfseries]
  \item[E1 Effect truth.]
  The issuer's claim satisfies $\ProducerSound$; authentication proves origin,
  not semantic truth.
  \item[E2 Base-model adequacy.]
  One selected root, sequential execution, static layering, a stratified
  acyclic chain, and no untracked cross-layer aliases.
  \item[E3 Alias admission.]
  The registered C \texttt{restrict}/LLVM \texttt{noalias} fact and the
  case-specific separation needed by the observation.
  \item[E4 Authentication and measurement.]
  HMAC, hashing, parsing, snapshots, filesystem measurement, and receipt
  checking correctly implement the abstract boolean gates.
  \item[E5 Compiler/model correspondence.]
  The registered LLVM root, load, call, callback operand, or GEP corresponds to
  the abstract request and observation.
  \item[E6 Selected-body adequacy.]
  Linker ownership and the runtime call target select the receipt-bound
  provider body.  This remains a premise for the dynamic-library routes.  The
  static profile checks it for one embedded-provider call edge while trusting
  its link-map/tool parsers, disassembly classification, and linker.
  \item[E7 Trusted-entry non-bypass.]
  The driver, internal pass API, plugin, key, and link gate form the stated
  trusted computing base.
  \item[E8 Callback adequacy.]
  There is one direct sequential callback; the provider truthfully forwards
  the declared same-root/order argument and both component effects are true.
  \item[E9 Projection adequacy.]
  The selected-slot \textsf{Stable} assertion is true, and ABI, layout, offset,
  width, address spaces, and GEP normalization agree with the model.
  \item[E10 Checked-producer adequacy.]
  The concrete Toka AST and emitted root summary correspond to the bounded
  source-command model.  Separately, LLVM parsing and the C++ checker's
  classification of the selected function correspond to the bounded
  concrete-IR action list.  The accepted exact IR is mechanically fixed as the
  registered backend input, but checker-to-model correspondence, source-to-IR
  compiler correctness, and backend correctness remain external premises.
\end{description}

\subsection{One complete projection trace}
\label{app:cairo-trace}

The Cairo case illustrates how the three proof layers meet.  Neutral caller IR
loads the pointer-valued \texttt{cairo\_path\_t.data} slot at arm64 byte offset
8, calls opaque \texttt{cairo\_append\_path}, and reloads that same slot.
The detached artifact asserts \textsf{Stable} for the exact projection
identity $(\mathit{layout},8,8,0,0)$ and the registered root argument.  The
validator checks the artifact origin and provider/ABI/layout identity; the
receipt then binds the neutral IR, canonical claim, provider snapshot,
sidecar, plugin, toolchain, commands, and caller object.  Only after all
current inputs equal those receipt claims does the internal pass receive the
selected-slot grant.  Its cache key matches the pre-call load, so it retains
the call and GEP but replaces the second load with the cached pointer.  The
runtime observation is unchanged.

A synthetic pointer-valued sibling at offset 16 has the same root and layout
but a different projection identity.  It cannot satisfy the request-identity
equality above, so no grant is issued and both loads remain.  This trace
demonstrates authorization confinement; it does not infer Cairo's effect or
prove general aggregate-path analysis.

\subsection{Complete checked-producer traces}
\label{app:toka-producer-trace}

Each accepted Toka input contains one public function,
\texttt{mutate\_inner}, whose selected root is used only for bounded topology
traversal and descendant payload access.  The compiler first emits a
preliminary native object and an object-bound root summary.  The source gate
checks the frozen compiler identity, source/evidence/object bindings, function
and root selection, empty function effects, and the
\{\textsf{Read},\textsf{Write}\} root-effect whitelist.

The same registered compiler separately emits exact unoptimized LLVM~20 IR.
The exact-IR checker scans the complete selected function
under a request fixing the function, root argument and order, target triple,
DataLayout, and IR digest.  The certificate and producer receipt separately
bind checker identity.
The three frozen positives certify,
respectively, traversal/read/write counts $(2,1,1)$ for the stable update,
$(2,1,0)$ for the read-only provider, and $(4,2,2)$ for the multi-payload
provider.  Only those accepted IR bytes are passed to the fixed
\texttt{llc} invocation.  Its output is a distinct final object, and the
ordinary order-two $G(0,1)$ artifact binds that final object rather than the
preliminary one.  For all three providers the consumer retains the call,
changes the caller from two root loads to one, and native execution returns
34, 28, and 46; removing caller \texttt{noalias} leaves both loads.

The rejection matrix exercises 14 concrete-IR classes: topology store,
partial-width topology store, scalar load from a topology slot, unknown direct
call, indirect call, pointer return, \texttt{ptrtoint}, \texttt{inttoptr},
dynamic GEP, pointer \texttt{phi}, pointer \texttt{select}, atomic load,
volatile store, and unsupported opcode.  Malformed IR is a tool error rather
than acceptance, and forged metadata cannot authorize a rejected topology
store.  The source-summary gate rejects four summaries: handle rebind,
root consumption/invalidation, raw pointer use, and unsafe code.

A second deterministic corpus is independent of Toka lowering.  LLVM's
verifier accepts all 17 modules before classification.  Six expected-preserving
modules vary conditional control flow, scalar payload type, root spills, and
read/write mode; 11 LLVM-valid mutations exercise exact rejection reasons for
topology writes, atomic/volatile access, pointer casts and comparisons,
non-\texttt{inbounds}/dynamic GEPs, calls, pointer merging, and escape.  All
decisions match the fixed manifest.  This corpus is bounded regression
coverage, not random fuzzing, exhaustive coverage, or a checker-correctness
argument.

\begin{table}[t]
\centering
\small
\caption{Warm-host checked-producer measurements on the registered
arm64 macOS/LLVM~20 host.  Both panels use three discarded warmups and 21
samples with \texttt{perf\_counter\_ns}; P95 is nearest rank.  Tool discovery
and checker/plugin construction are excluded.  Panel~(a) includes process
startup and file I/O within each named phase.  Panel~(b) times a prebuilt
checker subprocess, including IR read/parse/verify/classify and certificate
serialization; IR generation, file creation, and preflight are excluded.
These are descriptive prototype observations, not scaling results.}
\label{tab:checked-producer-cost}
\begin{tabular}{@{}lrr@{}}
\toprule
\multicolumn{3}{c}{(a) One accepted producer construction (ms)} \\
\cmidrule(lr){1-3}
Phase & Median & P95 \\
\midrule
Preliminary object + \texttt{.tke} & 243.866 & 245.841 \\
Source-summary gate & 0.385 & 0.430 \\
Exact LLVM IR emission & 118.825 & 121.518 \\
Exact-IR gate & 18.984 & 19.440 \\
Fixed \texttt{llc} & 44.275 & 44.686 \\
Sidecar/receipt issuance & 106.916 & 107.796 \\
\addlinespace
Construction total & 533.559 & 535.356 \\
Receipt revalidation & 127.138 & 129.871 \\
Construction + revalidation & 660.697 & 663.790 \\
\midrule
\multicolumn{3}{c}{(b) Accepted-IR checker response} \\
\cmidrule(lr){1-3}
Instructions / textual IR & Median (ms) & P95 (ms) \\
\midrule
8 / 0.7\,KiB & 10.879 & 11.039 \\
29 / 2.0\,KiB & 10.740 & 10.999 \\
113 / 7.1\,KiB & 10.946 & 11.205 \\
449 / 27.6\,KiB & 11.275 & 11.568 \\
1,793 / 109.8\,KiB & 12.576 & 12.983 \\
7,169 / 438.3\,KiB & 17.764 & 18.270 \\
\bottomrule
\end{tabular}
\end{table}

Finally, the producer receipt binds the source, compiler, preliminary object,
summary evidence, exact IR, checker, checker certificate, backend executable,
backend flags, final object, and sidecar.  Mutation of any one of those
materials, or of the receipt itself, fails validation.  These traces replace a
manual effect assertion only in the bounded checked-producer route.  They are
synthetic and do not prove the checker-to-model, compiler, or backend adequacy
premises in E10, nor do they supply an unchanged real-client positive.

\subsection{Pinned Linux/ARM64 functional replay}

The checked-producer route was freshly rebuilt inside a pinned
Ubuntu image for Linux on ARM64 with LLVM~20.
Table~\ref{tab:linux-replay} reports the
complete functional result; each positive uses newly emitted exact IR, an ELF
provider object, and a fresh Linux receipt rather than replaying Darwin
authorization material.

\begin{table}[h]
\centering
\small
\caption{Linux/ARM64 checked-producer replay.  Every positive retains one
opaque call.}
\label{tab:linux-replay}
\begin{tabular}{@{}lrrrr@{}}
\toprule
Case & Reads & Writes & Loads & Runtime \\
\midrule
stable & 1 & 1 & $2\!\rightarrow\!1$ & 34 \\
readonly & 1 & 0 & $2\!\rightarrow\!1$ & 28 \\
multi & 2 & 2 & $2\!\rightarrow\!1$ & 46 \\
\midrule
\multicolumn{5}{@{}l@{}}{Reject: 4 source; 14 IR; 11 tamper; 4 Darwin material.} \\
\bottomrule
\end{tabular}
\end{table}

The evidence binds the base-image digest, direct package inventory and built
image identity, Toka commit/compiler, LLVM checker, fixed backend, target
triple and DataLayout, and final objects.  The no-\texttt{noalias} control
retains two loads.  Because the host supplies the LinuxKit kernel, its release
is recorded but not treated as pinned.  No timing was collected, and this
replay makes no claim about bare-metal Linux, arbitrary Linux libraries,
cross-platform receipt reuse, or whole-compiler correctness.

\subsection{Closest-mechanism lowering audit}

The executable invariance control begins from the same Cairo neutral module,
canonical effect, provider snapshot, and non-consumer receipt claims as the
custom projection route.  Because a receipt binds the optimizer entry, its
consumer schema is necessarily distinct.  After validation, the control
introduces a fresh \texttt{invariant.group} identity only for one exact
selected-load/call/selected-load window, invokes stock GVN, and strips every
temporary assumption.  The frozen matrix records custom and invariance
$2\!\rightarrow\!1$ parity, a two-load external-\texttt{noescape} control,
selected/sibling separation, and rejection of injected invariance or
receipt/effect mismatches.  This is executable evidence about a closest
lowering target; no Rocq theorem claims equivalence of the two LLVM
implementations.  The external annotation is an independent user-responsible
claim, not something derived from the selected-slot receipt.

\subsection{Static-target certificate audit}
\label{app:static-target}

The Darwin/arm64 profile links only privately staged caller and provider object
bytes and passes no user-supplied link arguments.  Its target checker requires:
(1) the raw map assigns \texttt{g1\_loop} to
\texttt{verified-caller.o} and \texttt{mutate\_inner} to
\texttt{verified-provider-snapshot.o}; (2) the final Mach-O is an ARM64
executable with unique strong text definitions; (3) no exact protected symbol
occurs in bind, lazy-bind, weak-bind, chained-fixup import, or indirect-symbol
output; and (4) exactly one caller \texttt{BL} decodes to the final callee
address.  The certificate records these facts and the fixed LLVM-tool
identities.  A second receipt authenticates the certificate plus the prior
receipt, caller/provider digests, image, raw map, symbols, environment, and
profile.

The frozen stable, read-only, and multi-write images have zero registered
protected imports, preserve the call and one-load consumer, and return 34, 28,
and 46.  Ten material mutations, link-map ownership reassignment, a dynamic
provider, and additional link arguments reject.  A same-name dylib injection
loads its constructor but cannot redirect the protected direct branch.  These
are executable checks of one registered final-image shape, not a formal
semantics for Mach-O, AArch64, or the linker.

\subsection{Transaction-timing audit}

The frozen cost record contains 21 paired samples after three discarded
warmups and alternates stock-first with authorized-first order.  Each stock
sample times caller/main code generation and native linking.  Each authorized
sample times authorization and snapshotting, optimize/recheck/codegen, and
final receipt recheck, private snapshot staging, and linking.  Plugin
construction, neutral C-to-IR generation, sidecar issuance, and the separately
checked runtime oracle are not timed.  The checker recomputes median,
nearest-rank P95, minimum, and maximum from every raw phase, checks each
per-pair delta, and binds the manuscript's rounded totals.  This evidence is
descriptive for the registered microfixture and has no associated cost
theorem.

\subsection{Bounded IR-level applicability audit}
\label{app:ir-search}

As a second applicability check, we compiled four unchanged Rust FFI
snapshots to optimized LLVM IR and searched only for the complete
straight-line fragment accepted by the projection consumer: a simple pointer
load from a constant, \texttt{inbounds}, nonzero projection of a
\texttt{noalias} argument; a direct external
\texttt{void(ptr,ptr)} C call receiving that argument; and a second pointer
load from the identical projection.  Provider effects were to be audited only
after this syntactic filter produced a candidate.  Table~\ref{tab:ir-search}
reports the frozen result.

\begin{table}[h]
\centering
\scriptsize
\setlength{\tabcolsep}{3pt}
\caption{Bounded optimized-IR search.  ``Proj.'' counts admissible nonzero
constant projections; ``ABI+NA'' counts frozen-ABI calls receiving a
\texttt{noalias} argument.}
\label{tab:ir-search}
\begin{tabular}{@{}lrrrrrr@{}}
\toprule
Snapshot & Funcs & Calls & Ptr loads & Proj. & ABI+NA & Windows \\
\midrule
freetype-rs 0.38.0 & 53 & 201 & 57 & 24 & 0 & 0 \\
rust-harfbuzz 0.7.0 & 48 & 127 & 38 & 0 & 0 & 0 \\
tree-sitter 0.27.0 & 228 & 1,626 & 404 & 91 & 12 & 0 \\
cubeb-rs \texttt{tone} & 33 & 215 & 79 & 26 & 1 & 0 \\
\midrule
Total & 362 & 2,169 & 578 & 141 & 13 & 0 \\
\bottomrule
\end{tabular}
\end{table}

The reporter used LLVM~22.1.8 because the frozen Rust compiler emitted
LLVM~22 IR, whereas the evaluated consumer uses LLVM~20.1.8.  To check this
version boundary, both implementations consumed one identical frozen positive
module.  The reporter found exactly one window, and the authorized LLVM~20
consumer retained the call while reducing the selected pointer loads from two
to one.  This is a positive alignment control, not a proof that the reporter
and consumer recognize identical IR sets.

The four-snapshot result is neither an ecosystem estimate nor evidence that no
unchanged client exists.  It strengthens only the explanation of the observed
zero: many individual prerequisites occur in optimized IR, but their complete
conjunction did not occur in this bounded corpus.  The search did not change
the consumer, effect language, alias admission, ABI descriptor, or proof
model.

\fi
\end{document}